\documentclass[10pt,twocolumn,letterpaper]{article}

\usepackage{iccv}
\usepackage{times}
\usepackage{epsfig}
\usepackage{graphicx}
\usepackage{amsmath}
\usepackage{amssymb}
\usepackage{multirow}
\usepackage{arydshln}
\usepackage{booktabs}
\usepackage{mathtools}
\usepackage{caption}
\usepackage{subcaption}
\usepackage[pagebackref=true,breaklinks=true,letterpaper=true,colorlinks,bookmarks=false]{hyperref}

 \iccvfinalcopy 

\begin{document}

\title{AdaDM: Enabling Normalization for Image Super-Resolution}

\author{Jie Liu~~~~~~~~Jie Tang\thanks{Corresponding author.}~~~~~~~~Gangshan Wu\\
State Key Laboratory for Novel Software Technology, Nanjing University, China\\
{\tt\small jieliu@smail.nju.edu.cn, \{tangjie,gswu\}@nju.edu.cn}\\
{\small\url{https://github.com/njulj/AdaDM}}
}

\maketitle
\ificcvfinal\thispagestyle{empty}\fi

\begin{abstract}
    Normalization like Batch Normalization (BN) is a milestone technique to normalize the distributions of intermediate layers in deep learning, enabling faster training and better generalization accuracy. However, in fidelity image Super-Resolution (SR), it is believed that normalization layers get rid of range flexibility by normalizing the features and they are simply removed from modern SR networks. In this paper, we study this phenomenon quantitatively and qualitatively. We found that the standard deviation of the residual feature shrinks a lot after normalization layers, which causes the performance degradation in SR networks. Standard deviation reflects the amount of variation of pixel values. When the variation becomes smaller, the edges will become less discriminative for the network to resolve. To address this problem, we propose an Adaptive Deviation Modulator (AdaDM), in which a modulation factor is adaptively predicted to amplify the pixel deviation. For better generalization performance, we apply BN in state-of-the-art SR networks with the proposed AdaDM. Meanwhile, the deviation amplification strategy in AdaDM makes the edge information in the feature more distinguishable. As a consequence, SR networks with BN and our AdaDM can get substantial performance improvements on benchmark datasets. Extensive experiments have been conducted to show the effectiveness of our method. 
\end{abstract}

\section{Introduction}
Image Super-Resolution (SR) serves as a fundamental tool in image processing and computer vision. The goal of classical image SR is to reconstruct 
a high-fidelity High-Resolution (HR) image from a degraded Low-Resolution (LR) image. Since the pioneering work of SRCNN~\cite{SRCNN}, Convolutional
Neural Networks (CNN) have become the primary workhorse for image SR~\cite{FSRCNN,EDSR,IMDN,CARN,DBLP:conf/eccv/LeeLKH20,DRN,Wang_2021_ICCV}.
\begin{figure}
    \centering
    \begin{subfigure}[b]{\linewidth}
        \centering
        \includegraphics[width=\linewidth]{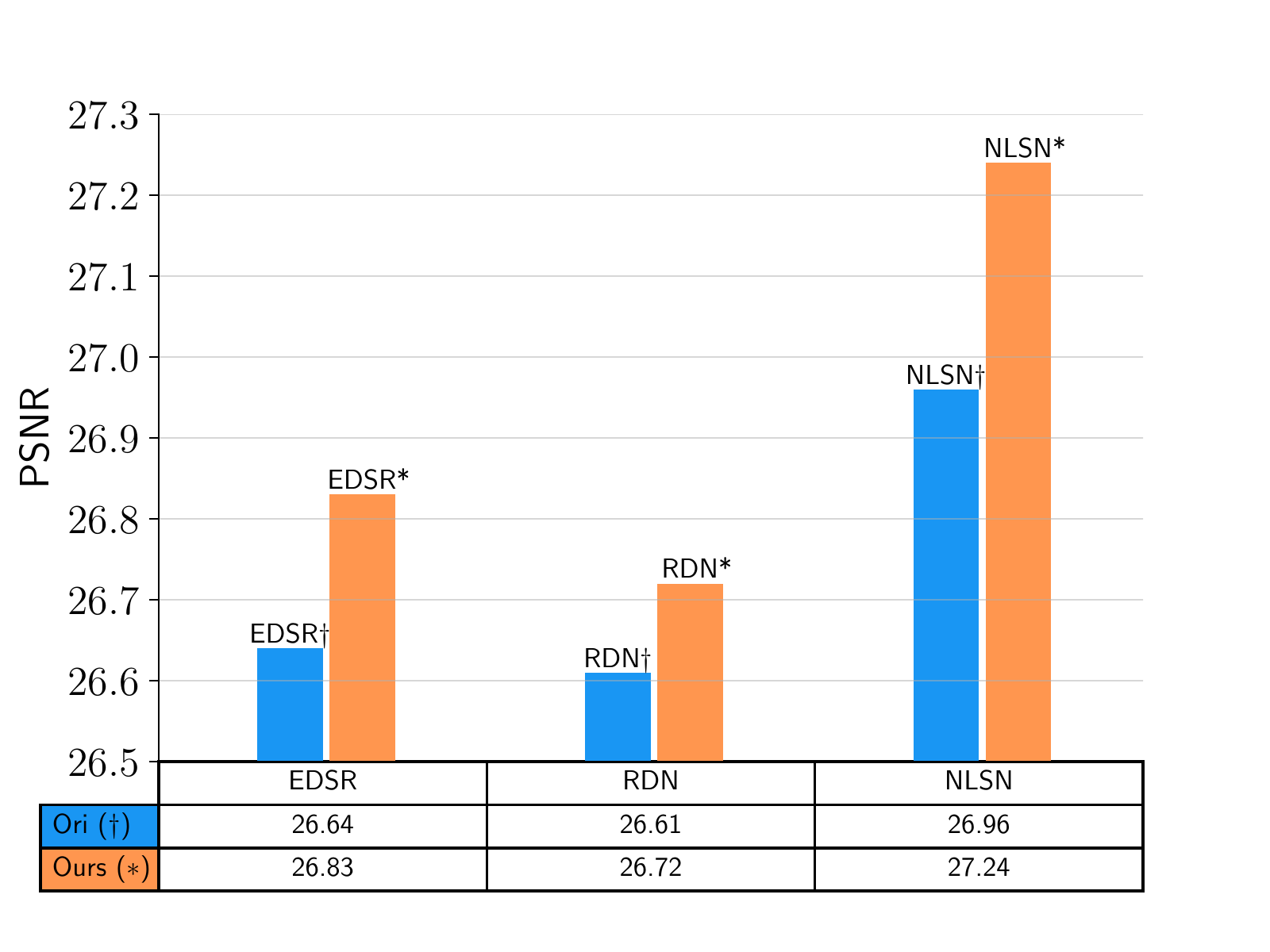}
        \caption{PSNR Improvements.}\label{fig:psnr_bar}
    \end{subfigure}
    \begin{subfigure}[b]{0.23\linewidth}
        \centering
        \includegraphics[width=\linewidth]{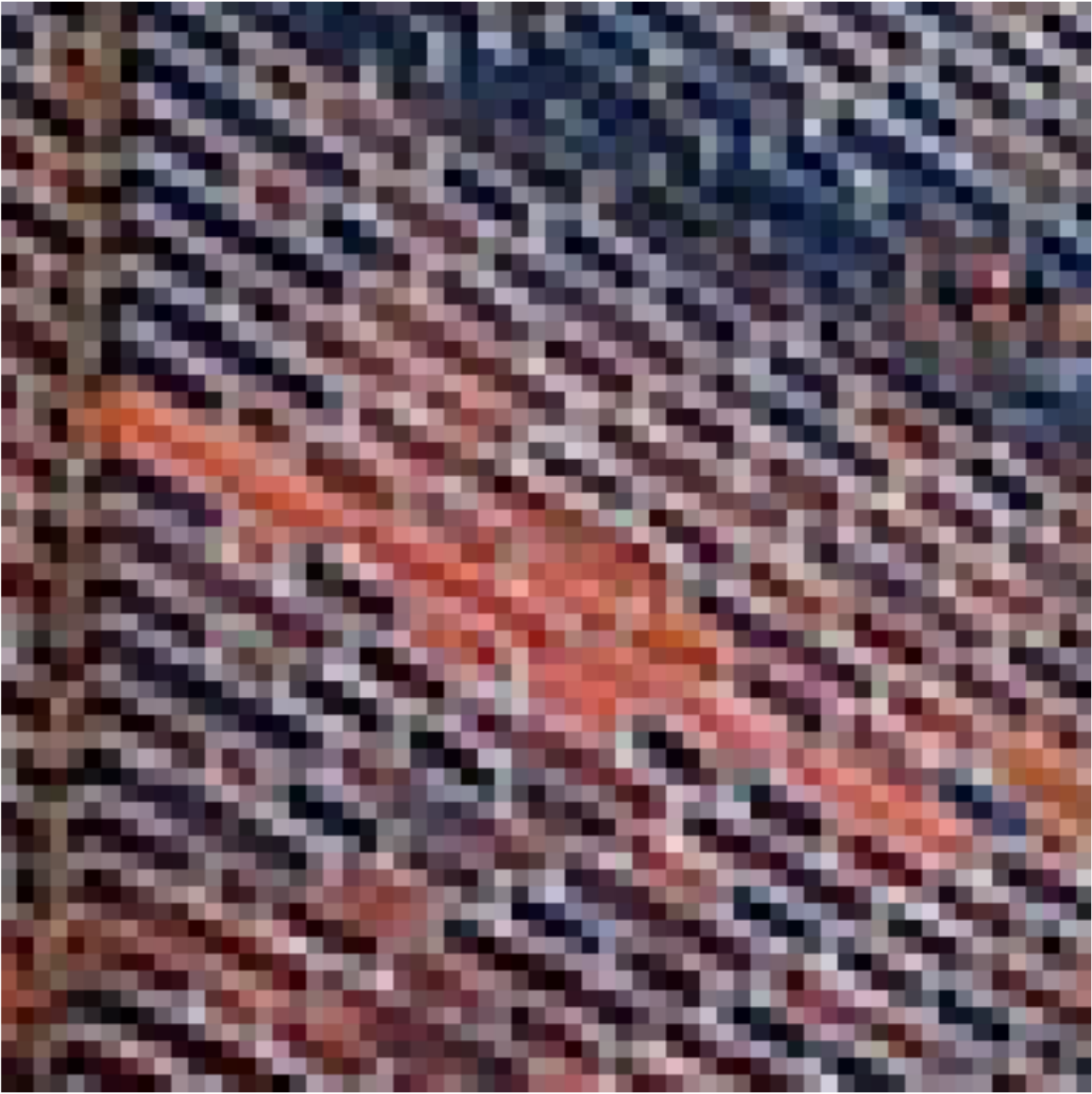}
        \caption{HR}\label{fig:demo_HR}
    \end{subfigure}
    \hspace{-4pt}
    \begin{subfigure}[b]{0.23\linewidth}
        \centering
        \includegraphics[width=\linewidth]{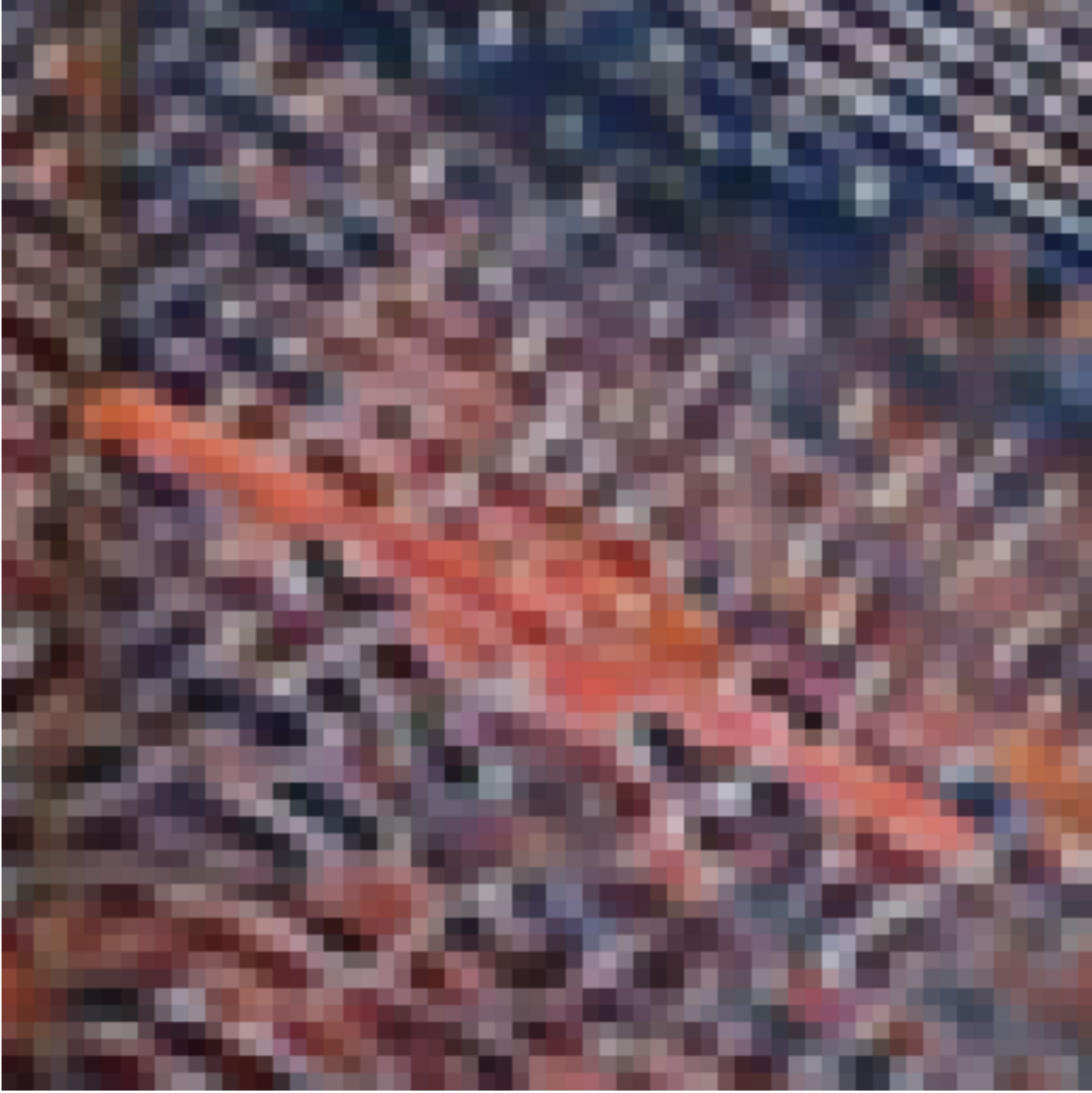}
        \caption{EDSR}\label{fig:demo_EDSR}
    \end{subfigure}
    \hspace{-4pt}
    \begin{subfigure}[b]{0.23\linewidth}
        \centering
        \includegraphics[width=\linewidth]{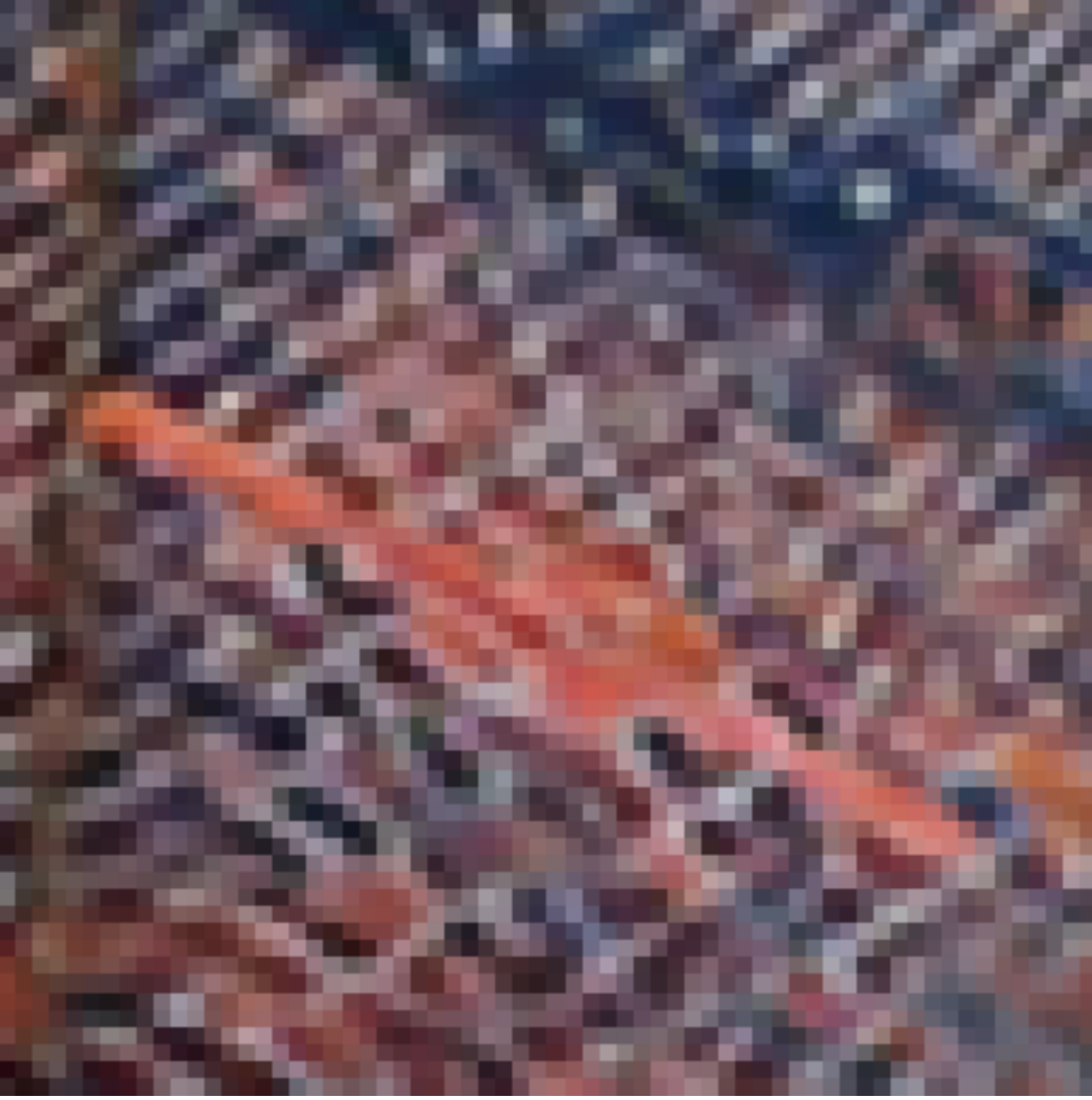}
        \caption{w/ BN}\label{fig:demo_BN}
    \end{subfigure}
    \hspace{-4pt}
    \begin{subfigure}[b]{0.23\linewidth}
        \centering
        \includegraphics[width=\linewidth]{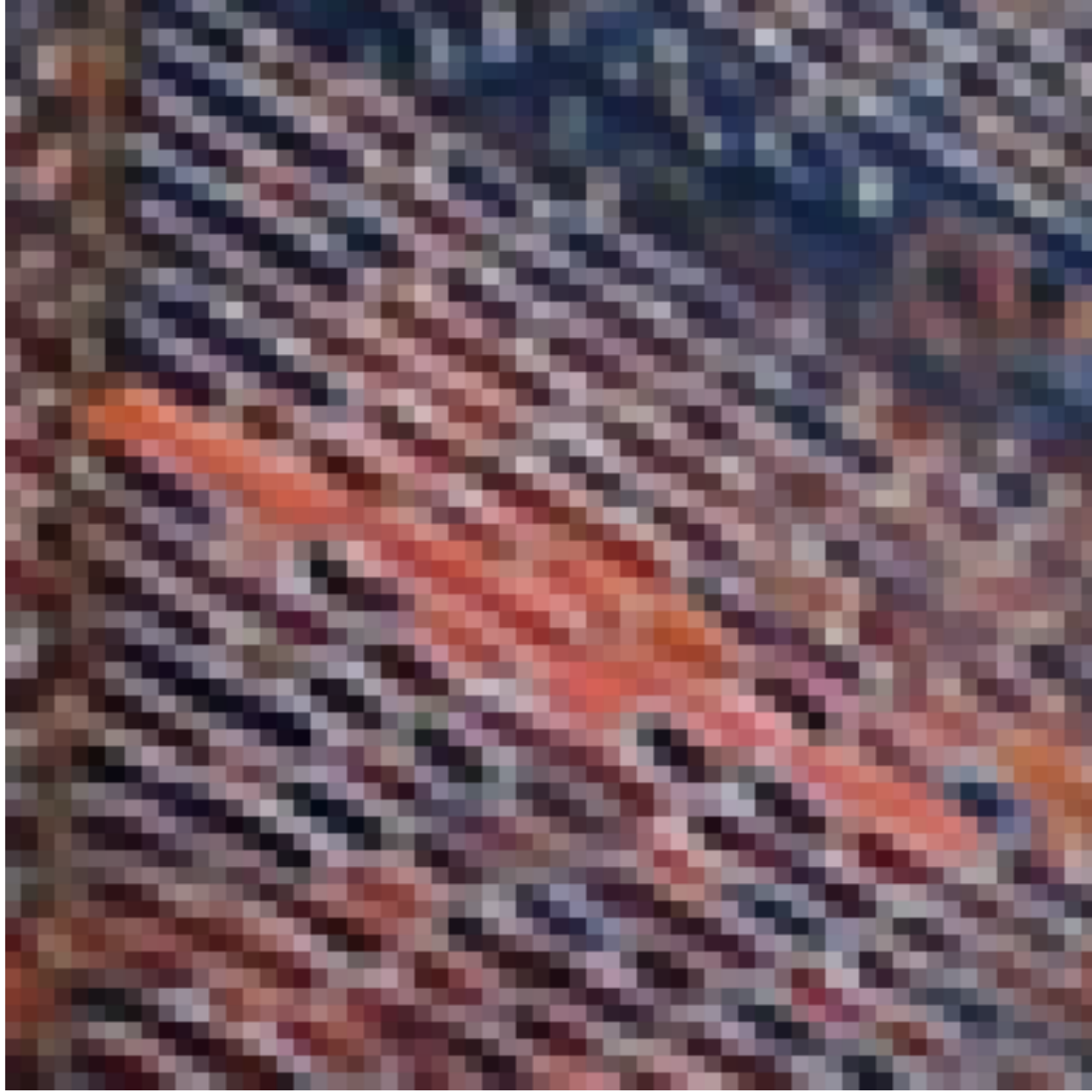}
        \caption{EDSR*}\label{fig:demo_AdaDM}
    \end{subfigure}
    \caption{(a) PSNR improvements of EDSR~\cite{EDSR}, RDN~\cite{RDN} and NLSN~\cite{NLSN} with BN+AdaDM on Urban100 $\times 4$ dataset.
        (b) HR patch from img098 in Urban100 dataset. (c) SR results of EDSR.
    (d) SR results of EDSR with BN layers. (e) SR results of EDSR with BN+AdaDM.
    }\label{fig:teaser}
\end{figure}

Early CNN-based SR networks, for example SRResNet~\cite{SRResNet}, applied the ResNet~\cite{ResNet} architecture to the super-resolution
task. However, Lim~\etal~\cite{EDSR} found that the Batch Normalization (BN) layers in residual blocks damage the performance of SR networks. 
They speculate that BN layers get rid of range flexibility from networks by normalizing the features. As a result, they removed
BN layers from the residual blocks and achieved substantial performance improvements. Inspired by this, most of the state-of-the-art 
SR networks exclude normalization layers in their building blocks. On the other hand, many SR networks are getting larger and larger in pursuit
of extreme fidelity. When the network becomes deeper and wider, it will be very difficult to optimize without normalization layers. Since normalization
layers have been successfully applied to various computer vision tasks for faster training and better generalization, we wonder whether they can be
applied in SR networks without compromising performance. 

In this paper, we show that normalization will reduce 
the standard deviation of feature pixels (\ie, pixel deviation), which causes the performance degradation. Standard deviation reflects the amount
of variation of pixel values. As the pixel deviation shrinks, the edge information in the feature becomes less discriminative. Figure~\ref{fig:demo_BN} shows 
one sample output of the EDSR model with BN layers. As we can see, it looks more blurry and messy compared to the patch produced by original
EDSR in Figure~\ref{fig:demo_EDSR}, which indicates that the BN layers increase the difficulty of the network to resolve the edges. To address this 
problem, we propose an Adaptive Deviation Modulator (AdaDM) to amplify the pixel deviation of the features in cooperation with normalization
layers. Figure~\ref{fig:demo_AdaDM} shows the SR patch produced by the EDSR model with BN layers and our AdaDM. It reconstructs the edges 
sharper and clearer than EDSR and EDSR with BN, which demonstrates the effectiveness of our AdaDM.

With the proposed AdaDM, we can successfully apply normalization layers in state-of-the-art SR networks without compromising performance.
We insert BN layers and our AdaDM into the residual blocks of EDSR~\cite{EDSR}, RDN~\cite{RDN} and NLSN~\cite{NLSN} to build three new 
models: EDSR*, RDN* and NLSN*. As shown in
Figure~\ref{fig:psnr_bar}, the performance of the three models has been significantly improved since they can benefit from feature normalization,
which shows the necessity of normalization in large models. 
Overall, the main contribution of this paper is to propose a novel AdaDM that can enable feature normalization in SR networks, which greatly improves the performance for high-fidelity SR. Code and pretrained models are available at \url{https://github.com/njulj/AdaDM}.

\section{Related Work}
\subsection{Image Super-Resolution}
In the past few years, numerous SR methods based on deep learning have been proposed~\cite{FSRCNN,LapSRN,ESPCN,MemNet,SRResNet,DBPN,SRMD,SRFBN,MRNet,OISR,DBLP:conf/cvpr/MeiFZHHS20,RFDN,ESRGAN,AdderSR}. The pioneering work was done by~\cite{SRCNN}, who first proposed a three-layer
convolutional network (SRCNN) to learn an end-to-end mapping from LR to HR directly.
Later, VDSR~\cite{VDSR} and DRCN~\cite{DRCN} increase the network depth to 20 and achieve notable improvements over SRCNN. EDSR~\cite{EDSR} further
increases the network depth to about 165 layers by removing BN layers from the residual blocks~\cite{ResNet}. ~\cite{RDN} proposed to use dense 
connections~\cite{DenseNet} in a residual block to get a better performance. RCAN~\cite{RCAN} uses the residual in residual (RIR) connections to
train a very deep network that achieves excellent performance. Recently, many attention based SR 
networks~\cite{NLRN,NLRN,SAN,RFANet,HAN,NLSN,CRAN,DFSA,IPT,SwinIR} are proposed to exploit the extreme fidelity of image SR. 
However, most of them do not use normalization layers because of the poor performance.
Normalization like BN is a milestone technique in deep learning. It is a waste by simply abandoning BN in residual blocks.
In this paper, we show that the proposed AdaDM can successfully enable BN layers in SR networks, which greatly improves the performance.

\subsection{Feature Normalization}
Following~\cite{GN}, we give a general formulation of feature normalization in the case of 2D images. A series
of feature normalization techniques, including Batch Normalization (BN)~\cite{BN}, Layer Normalization (LN)~\cite{LN}, 
Instance Normalization (IN)~\cite{IN} and Group Normalization (GN)~\cite{GN}, normalize the features as follows:
\begin{equation}\label{eq:norm}
    \hat{x}_i = \frac{x_i - \mu_i}{\sigma_i},
\end{equation}
where $x$ is the intermediate feature of a network, and $i = (i_N, i_C, i_H, i_W)$ is a 4D vector indexing the
features in $(N, C, H, W)$ order. Here $N$, $C$, $H$ and $W$ denote the batch, channel, height and width dimensions,
respectively.

In (\ref{eq:norm}), $\mu$ and $\sigma$ are the mean and standard deviation computed as:
\begin{equation}\label{eq:mu_sigma}
    \mu_i = \frac{1}{m}\sum_{k\in S_i}x_k,\;\;\sigma_i = \sqrt{\frac{1}{m}\sum_{k\in S_i}(x_k-\mu_i)^2+\epsilon},
\end{equation}
with $\epsilon$ as a small constant for numerical stability. The mean and standard deviation are computed within a
set of pixels $S_i$, and $m$ is the size of $S_i$. For BN, $S_i$ is defined by:
\begin{equation}\label{eq:bn}
    S_i = \{ k\;|\;k_C = i_C \}.
\end{equation}
Here $k = (k_N, k_C, k_H, k_W)$, and $k_C$ is the sub-index of $k$ along the $C$ axis. This means that pixels in
the same channel are normalized together, \ie, $\mu$ and $\sigma$ are computed along the $(N, H, W)$ axes.
For LN, the set is:
\begin{equation}\label{eq:ln}
    S_i = \{ k\;|\;k_N = i_N \},
\end{equation}
indicating that $\mu$ and $\sigma$ are computed along the $(C, H, W)$ axes for each sample. The details for IN 
and GN can be found in~\cite{GN}.

\begin{figure*}[t]
    \centering

    \begin{subfigure}[b]{0.23\linewidth}
        \centering
    \includegraphics[width=\linewidth]{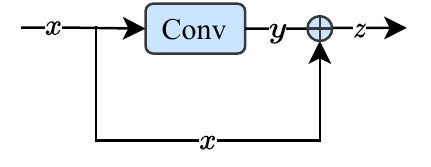}
    \caption{T1.}\label{fig:T1}
    \end{subfigure}
    \begin{subfigure}[b]{0.3\linewidth}
        \centering
    \includegraphics[width=\linewidth]{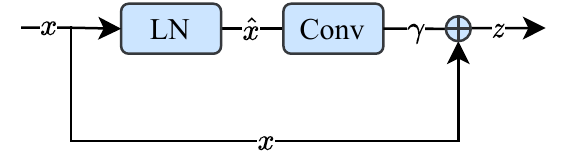}
    \caption{T2.}\label{fig:T2}
    \end{subfigure}
    \begin{subfigure}[b]{0.35\linewidth}
        \centering
    \includegraphics[width=\linewidth]{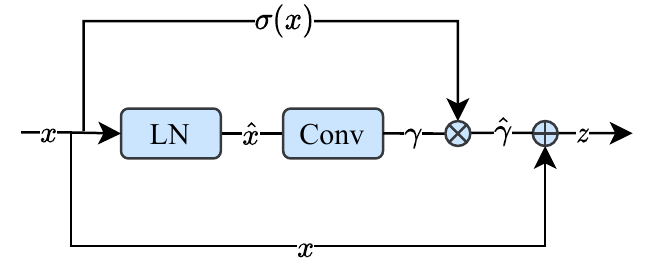}
    \caption{T3.}\label{fig:T3}
    \end{subfigure}
    \begin{subfigure}[b]{0.27\linewidth}
        \centering
    \includegraphics[width=\linewidth]{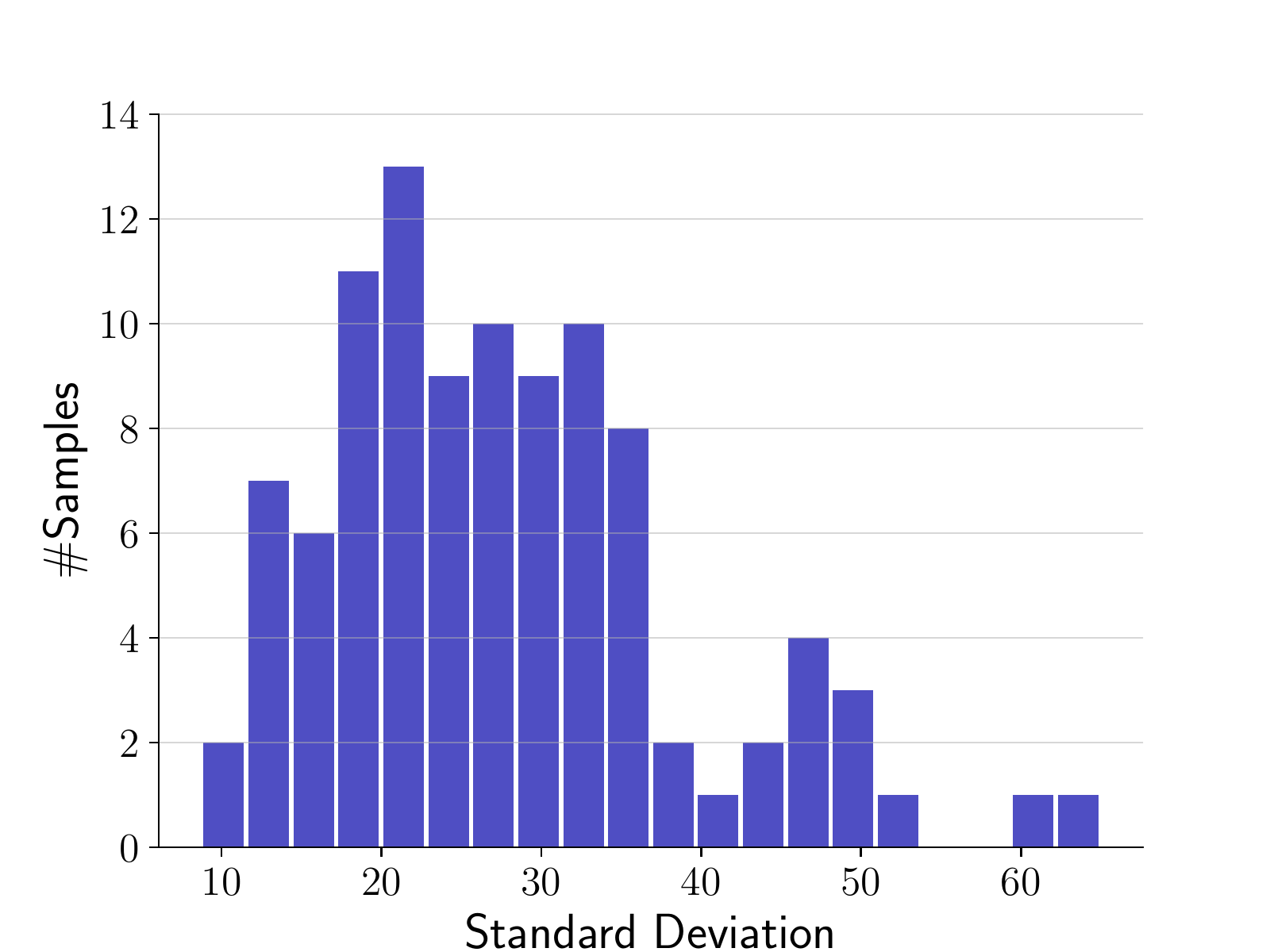}
    \caption{Deviation Histogram of T1.}\label{fig:T1_hist}
    \end{subfigure}
    \begin{subfigure}[b]{0.27\linewidth}
        \centering
    \includegraphics[width=\linewidth]{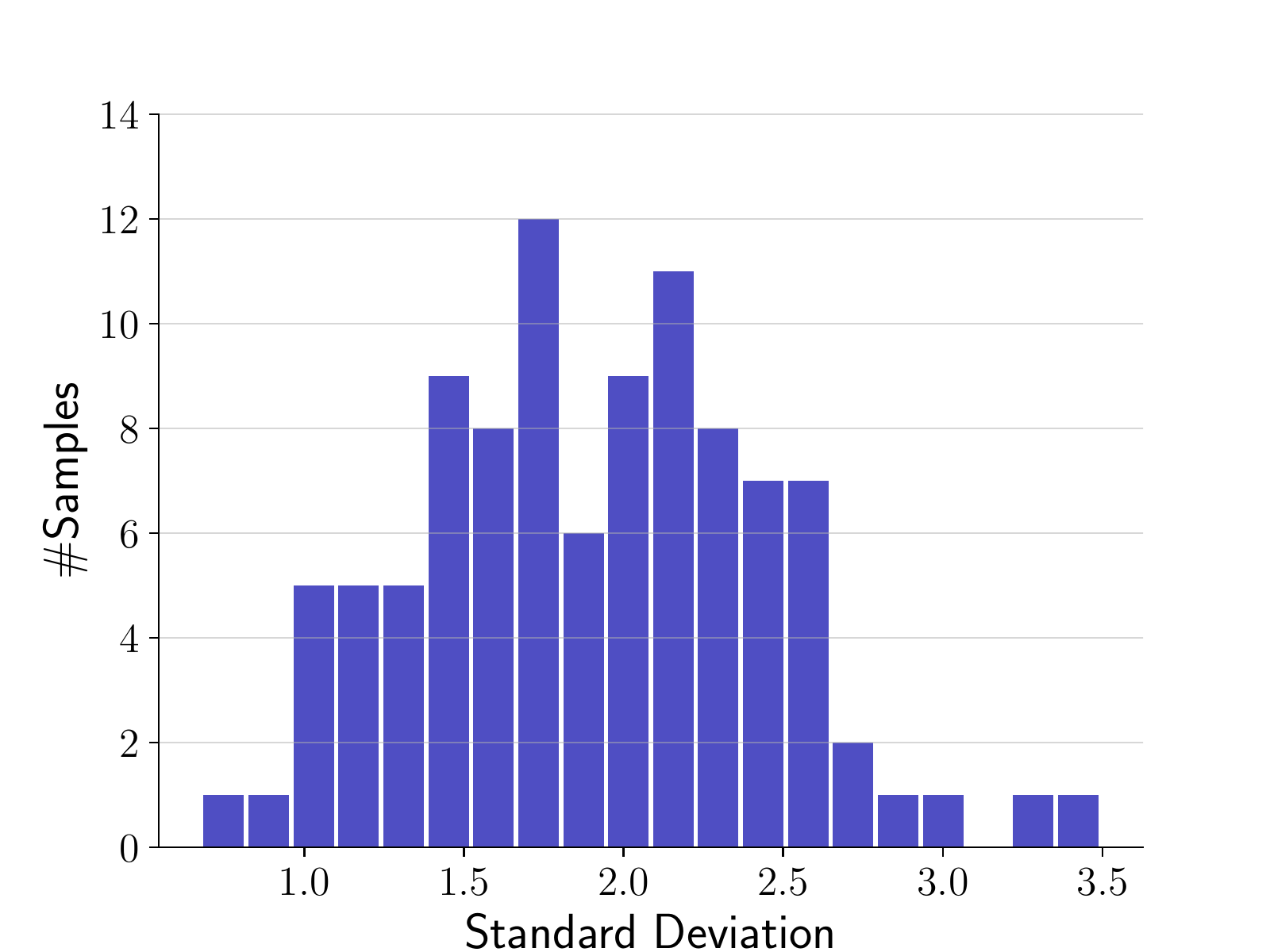}
    \caption{Deviation Histogram of T2.}\label{fig:T2_hist}
    \end{subfigure}
    \begin{subfigure}[b]{0.27\linewidth}
        \centering
    \includegraphics[width=\linewidth]{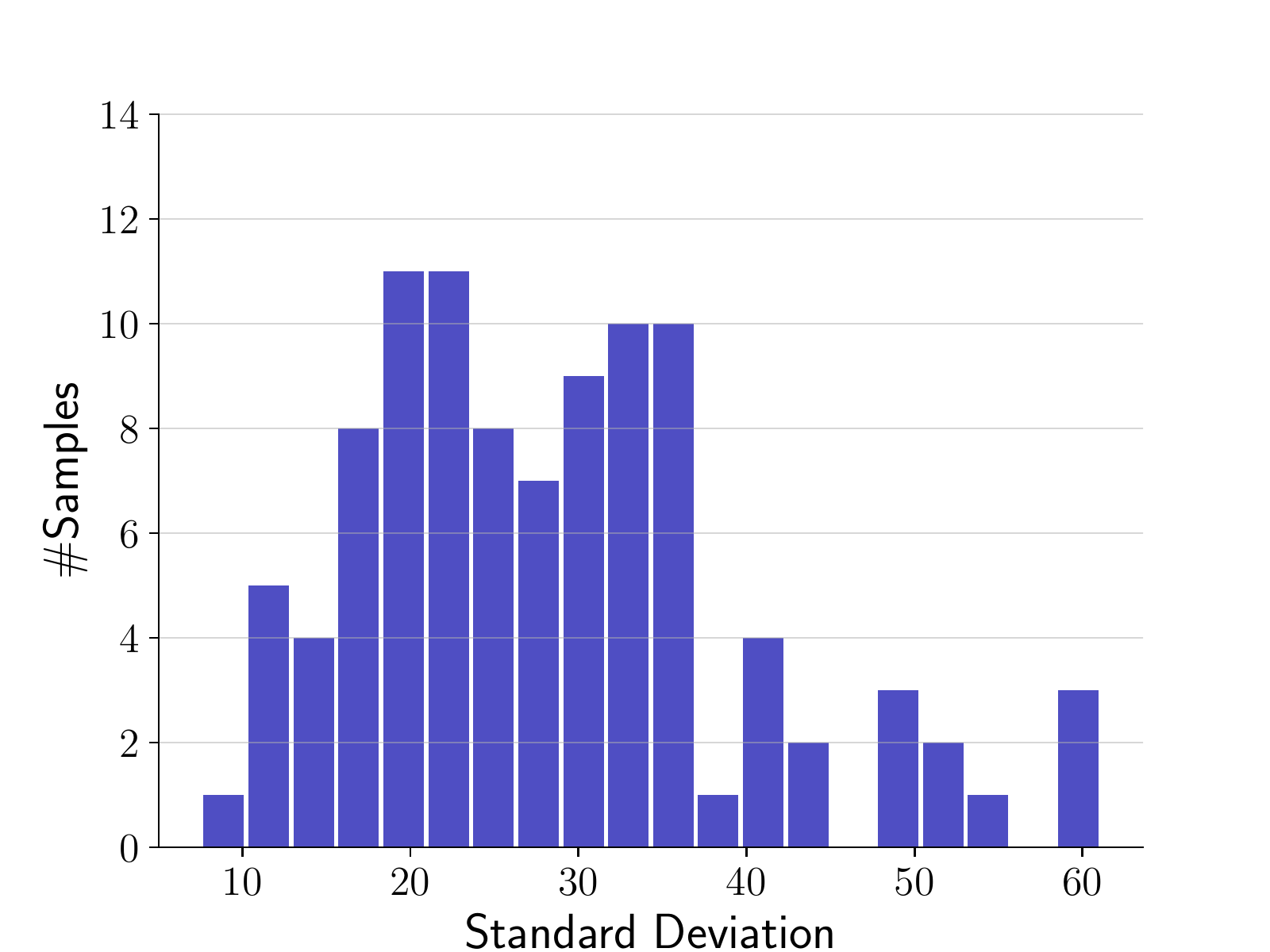}
    \caption{Deviation Histogram of T3.}\label{fig:T3_hist}
    \end{subfigure}
    \caption{(a)-(c): Three toy residual blocks. (d)-(f): Deviation histograms of $y$, $\gamma$ and $\hat{\gamma}$, respectively.
    The standard deviation is computed on the 100 images from Urban100 $\times 2$ dataset.}\label{fig:toy}
\end{figure*}
\begin{figure}
    \centering
    \includegraphics[width=0.8\linewidth]{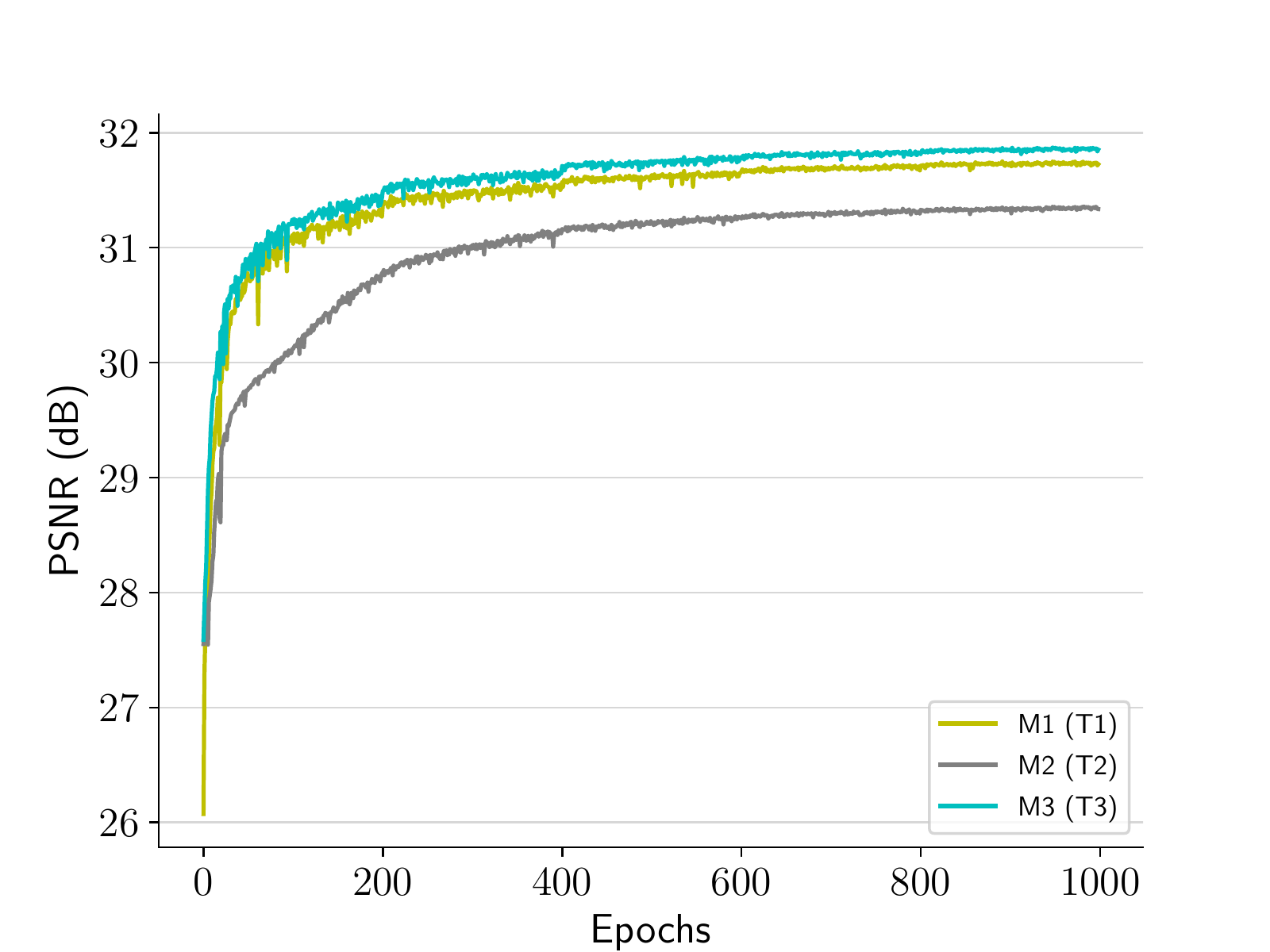}
    \caption{The learning curves of three toy SR models. Model M1/M2/M3 is constructed by using T1/T2/T3
    as the building block. All the models are evaluated on Urban100 $\times 2$ dataset.}\label{fig:toy_curve}
\end{figure}

\section{Motivation}\label{sec:motivation}
For a better understanding, we first construct three toy residual blocks, \ie, T1, T2 and T3, to illustrate the motivation behind our method. 
As shown in Figure~\ref{fig:toy}, T1 contains only one convolutional layer (Conv) in the residual branch. T2 is constructed 
by inserting a LN layer before the Conv layer. We use LN because it is easier to perform formal analysis than BN.
T3 has the same layers as T2 except for the new element-wise multiplication layer. Next we will give a formal description of T2
and then discuss the effects of feature
normalization from the perspective of pixel deviation. For simplicity, we remove the bias terms in Conv layers and
turn off the affine transformation in LN layers.

\begin{figure*}[t]
    \centering
    \begin{subfigure}[b]{0.15\linewidth}
        \centering
    \includegraphics[width=\linewidth]{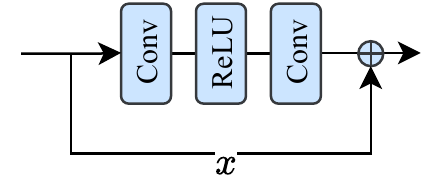}
    \caption{}\label{fig:RB}
    \end{subfigure}
    \begin{subfigure}[b]{0.24\linewidth}
        \centering
    \includegraphics[width=\linewidth]{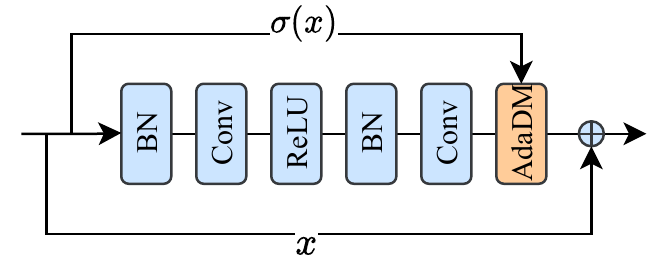}
    \caption{}\label{fig:RB_AdaDM}
    \end{subfigure}
    \begin{subfigure}[b]{0.25\linewidth}
        \centering
    \includegraphics[width=\linewidth]{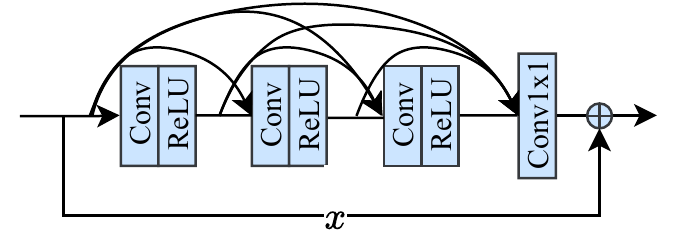}
    \caption{}\label{fig:RDB}
    \end{subfigure}
    \begin{subfigure}[b]{0.29\linewidth}
        \centering
    \includegraphics[width=\linewidth]{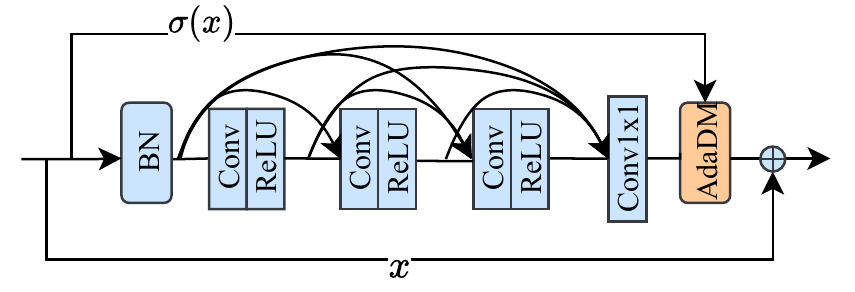}
    \caption{}\label{fig:RDB_AdaDM}
    \end{subfigure}
    \caption{(a) residual block in EDSR~\cite{EDSR} and NLSN~\cite{NLSN}. (b) residual block with AdaDM. 
    (c) residual dense block in RDN~\cite{RDN}. (d) residual dense block with AdaDM.}
\end{figure*}

It is worth noting that the following analysis is based on a forward propagation and we assume T1, T2 and T3
have the same predefined weights. 
In Figure~\ref{fig:T2}, $x$ is a single input sample with three axes $(C, H, W)$, and $\hat{x}$ is the transformed 
feature with LN. For LN, the mean and standard deviation are computed along the $(C, H, W)$ axes, so we can compute $\hat{x}$ as:
\begin{equation}\label{eq:T2_LN}
    \hat{x} = \frac{x - \mu}{\sigma},
\end{equation}
where $\mu$ and $\sigma$ are scalars shared by all pixels in $x$. Denote the function of the Conv layer as $f_{Conv}$, then the output $\gamma$ can be computed by:
\begin{equation}\label{eq:T2_gamma1}
    \gamma = f_{Conv}(\hat{x}) =  f_{Conv}(\frac{x - \mu}{\sigma}).
\end{equation}
According to the associativity of convolution to scalar multiplication, we can rewrite (\ref{eq:T2_gamma1}) as:
\begin{equation}
    \gamma = \frac{1}{\sigma}f_{Conv}(x - \mu).
\end{equation}
This formula can be further expanded using the distributivity:
\begin{equation}\label{eq:T2}
    \gamma = \frac{1}{\sigma}f_{Conv}(x) - \frac{1}{\sigma}f_{Conv}(\mu I).
\end{equation}
Here $I$ is the all-ones matrix with the same dimensions as $x$. In Figure~\ref{fig:T1}, the output of Conv layer 
can be written as:
\begin{equation}\label{eq:T1}
    y = f_{Conv}(x).
\end{equation}
Comparing (\ref{eq:T2}) and (\ref{eq:T1}), we can find that the pixel distributions (\eg mean and variation) of y
has been reshaped by LN. In image SR, we are more concerned about the pixel variation since it reflects some edge information in features.
Here we compute the standard deviation ($std$) of $\gamma$: 
\begin{align}
    std(\gamma) &= std(\frac{1}{\sigma}f_{Conv}(x) - \frac{1}{\sigma}f_{Conv}(\mu I)) \\
                &= \frac{1}{\sigma}std(f_{Conv}(x)) \\
                &= \frac{1}{\sigma}std(y).
\end{align}
As we can see, the pixel deviation is reduced to
$1/\sigma$ with feature normalization. The deviation is reduced because $\sigma$ is usually greater than 1
(Figure~\ref{fig:T1_hist}-\ref{fig:T3_hist}).
To compensate for the loss of pixel deviation, we multiply 
the $\gamma$ with $\sigma$ in T3. In Figure~\ref{fig:T3}, $\hat{\gamma}$ is obtained by:
\begin{equation}\label{eq:T2_DAM}
    \hat{\gamma} = \gamma\cdot \sigma,
\end{equation}
where $\sigma$ is the standard deviation of $x$. This process is referred to as Deviation Amplification (DA) in our
paper. As a result, $\hat{\gamma}$ can keep the original pixel deviation in the presence of LN, \ie,
$std(\hat{\gamma}) = \sigma std(\gamma) =  std(y)$.

We give the above idealized analysis to show the motivation of DA. Now, we will train three toy SR models, 
\ie, M1, M2 and M3, by using T1, T2 and T3 to show that the DA mechanism actually works very well in reality. We adopt
the EDSR~\cite{EDSR} network and replace the body part with 32 toy residual blocks. We use 64 feature
channels for fast model training. The learning curves are depicted in Figure~\ref{fig:toy_curve}. Among the three
models, M2 behaves much worse than M1, which is consistent with observations in~\cite{EDSR}.
The huge performance decline comes from the added LN layer that limits the pixel deviation of the residual branch.
M3 addresses this problem by using our DA mechanism and therefore achieves the best performance. Because of DA,
the Conv layer in M3 can benefit from feature normalization while its output feature can also keep a regular 
pixel deviation. In Figure~\ref{fig:T1_hist}-\ref{fig:T3_hist}, we plot the deviation histograms of T1-T3 on Urban100 dataset. 
The standard deviation is computed on $y$, $\gamma$ and $\hat{\gamma}$, respectively. It can be observed that 
T1 and T3 have similar deviation ranges while the deviation of T2 has been severely reduced, which is
consistent with our analysis.

\section{Method}
\subsection{Adaptive Deviation Modulator}\label{sec:AdaDM}
\begin{figure}
    \centering
    \includegraphics[width=0.4\linewidth]{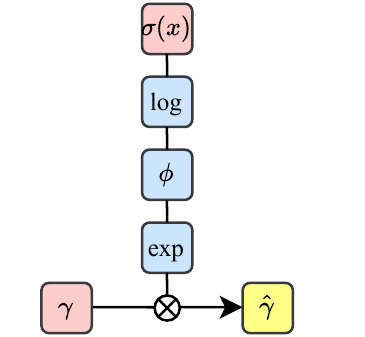}
    \vspace{6pt}
    \caption{Schema of AdaDM.}\label{fig:AdaDM}
\end{figure}
In this part, we turn to the SR models in reality. In real-world deep networks, the architecture will be much more
complicated. For example, there are bias terms in each convolution and there may be non-linear layers between 
convolutional layers. 
Moreover, BN is usually applied in convolutional neural networks. It keeps separate $\mu$ and $\sigma$
for each channel, which is different from LN. Though (\ref{eq:T2_LN})-(\ref{eq:T2_DAM}) may not hold for 
real-world cases, it gives us some insights for designing a workable solution for feature normalization. 
In general, we need a mechanism to recover the pixel deviation for the residual branch containing normalization
layers. In order to adapt to various network architectures, we propose an Adaptive Deviation Modulator (AdaDM)
, enabling the network to learn the deviation amplification during training. 
The process of AdaDM is depicted in Figure~\ref{fig:AdaDM} and it conducts the following computation: 
\begin{equation}\label{eq:AdaDM}
    \hat{\gamma} = \gamma\cdot e^{\phi(\log(\sigma(x)))}, 
\end{equation}
where $x$ is the input of a residual block, and $\sigma(x)$ computes the standard deviation of $x$ along the 
(C, H, W) axes (\ie, one $\sigma$ per 3D feature). $\gamma$ is the feature needed to be modulated and $\hat{\gamma}$ is the modulated output.
$\phi(v)\coloneqq w\cdot v + b$ is a learnable perceptron that consists of a weight $w$ and 
a bias $b$. During training, $w$ and $b$ can be updated via backpropagation algorithm. The function of $\phi$
is to predict a proper modulation factor according to the input scalar value $v$. In (\ref{eq:AdaDM}), we learn
$\phi$ in the logarithmic space (\ie, $v=\log(\sigma(x))$) for a better stability. The final modulation factor
is obtained by an exponential operation. In practice, $w$ is initialized to 1 and $b$ is initialized to 0 so
that (\ref{eq:AdaDM}) can be degenerated into (\ref{eq:T2_DAM}) by default. That is, the proposed AdaDM in
(\ref{eq:AdaDM}) applies a more general modulation strategy than the DA in (\ref{eq:T2_DAM}).

\noindent\textbf{Discussion.} For deviation amplification, one natural way is to compute the standard deviation and
predict a modulation factor for each channel. We do not adopt this method because assigning 
a scaling factor to each channel involves channel interdependencies. It is difficult to distinguish whether the 
contribution of this method comes from channel correlation modeling or deviation amplification or both. 
The purpose of this paper is to verify the effectiveness of our deviation amplification mechanism, so we choose to 
predict a single modulation factor for the whole feature.

%
\begin{figure}
    \centering
    \begin{subfigure}[b]{0.6\linewidth}
        \centering
    \includegraphics[width=\linewidth]{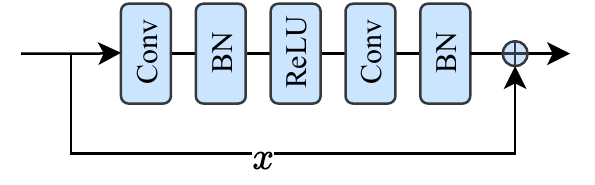}
    \caption{SRRB}\label{fig:SRResNet}
    \end{subfigure}
    \begin{subfigure}[b]{0.6\linewidth}
        \centering
    \includegraphics[width=\linewidth]{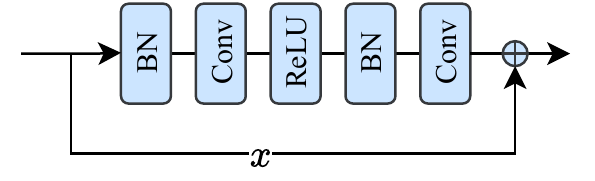}
    \caption{PreRB}\label{fig:AdaDM_BN}
    \end{subfigure}
    \caption{Two residual blocks used in Table~\ref{tab:RB_SRRB} and Table~\ref{tab:RB_SRRBPre}.}\label{fig:ablation_bn}
\end{figure}

\subsection{AdaDM with EDSR, RDN and NLSN}
For fidelity image SR, it is a common practice to avoid using normalization layers in the networks.
Now the network can benefit from normalization with our AdaDM.
In the following, we will show how to use normalization layers and our AdaDM collaboratively in 
state-of-the-art networks. Due to time and computing resource constraints, we consider three SR networks, \ie,
EDSR~\cite{EDSR}, RDN~\cite{RDN} and NLSN~\cite{NLSN}. We choose EDSR because it is widely used in image SR and its
simple architecture is very suitable for method verification. RDN is selected because of the dense connections which
are complicated enough to validate the effectiveness of our AdaDM. NLSN is one of the state-of-the-art SR models, which
is attention-based and has excellent performance.
EDSR and NLSN use the residual block in Figure~\ref{fig:RB}
while RDN uses the residual dense block in Figure~\ref{fig:RDB}. In Figure~\ref{fig:RB_AdaDM}, we add a BN layer before each
convolution and insert an AdaDM layer at the end of the residual branch to build a new residual block with feature normalization, 
which is used in EDSR* and NLSN*.
Figure~\ref{fig:RDB_AdaDM} depicts the new residual dense block, with a BN at the front and an AdaDM at the end. This block is used in RDN*.
In experiments, we will show that EDSR*, RDN* and NLSN* can achieve significant performance improvements, demonstrating the necessity of
normalization for better generalization performance.

\begin{table}
    \vspace{-1.2em}
    \centering
    \parbox{0.65\linewidth}{
        \caption{RB \vs SRRB}\label{tab:RB_SRRB}
    \resizebox{\linewidth}{!}{
    \begin{tabular}{ccc}
    \toprule
    Dataset & RB & SRRB \\
    \midrule
    Set5 & 38.22 / 0.9612 & 38.19 / 0.9612  \\
    Set14 & 33.97 / 0.9207 & 33.89 / 0.9200  \\
    B100 & 32.35 / 0.9019 & 32.32 / 0.9014 \\
    Urban100 & 32.96 / 0.9359 & 32.77 / 0.9343\\
    Manga109 & 39.13 / 0.9778 & 39.00 / 0.9778\\
    \bottomrule
    \end{tabular}}}
    \parbox{0.65\linewidth}{
        \caption{RB \vs PreRB.}\label{tab:RB_SRRBPre}
    \resizebox{\linewidth}{!}{
    \begin{tabular}{ccc}
    \toprule
    Dataset & RB & PreRB \\
    \midrule
    Set5 & 38.22 / 0.9612 & 38.21 / 0.9611  \\
    Set14 & 33.97 / 0.9207 &  33.97 / 0.9209 \\
    B100 & 32.35 / 0.9019 &  32.36 / 0.9020\\
    Urban100 & 32.96 / 0.9359 & 32.97 / 0.9360\\
    Manga109 & 39.13 / 0.9778 & 39.17 / 0.9782\\
    \bottomrule
    \end{tabular}}}
    \quad
    \parbox{0.9\linewidth}{
    \caption{Effects of AdaDM for $\times 2$ SR. ``--/--'' indicates that the model
        does not converge and there are no evaluation results. All the models are evaluated with PSNR / SSIM on five
    benchmark datasets.}\label{tab:ablation}
    \resizebox{\linewidth}{!}{
    \begin{tabular}{cccc}
    \toprule
    Dataset & EDSR & AdaDM only & EDSR* (full)\\
    \midrule
    Set5 & 38.22 / 0.9612 & -- / --  &  38.25 / 0.9615 \\
    Set14 & 33.97 / 0.9207 & -- / --  & 34.00 / 0.9205 \\
    B100 & 32.35 / 0.9019 & -- / --   & 32.37 / 0.9022\\
    Urban100 & 32.96 / 0.9359 & -- / --  &  33.12 / 0.9371\\
    Manga109 & 39.13 / 0.9778 & -- / --  & 39.31 / 0.9783 \\
    \bottomrule
    \end{tabular}}}
\end{table}

\begin{figure*}
    \centering
    \begin{subfigure}[b]{0.45\linewidth}
        \centering
        \includegraphics[width=\linewidth]{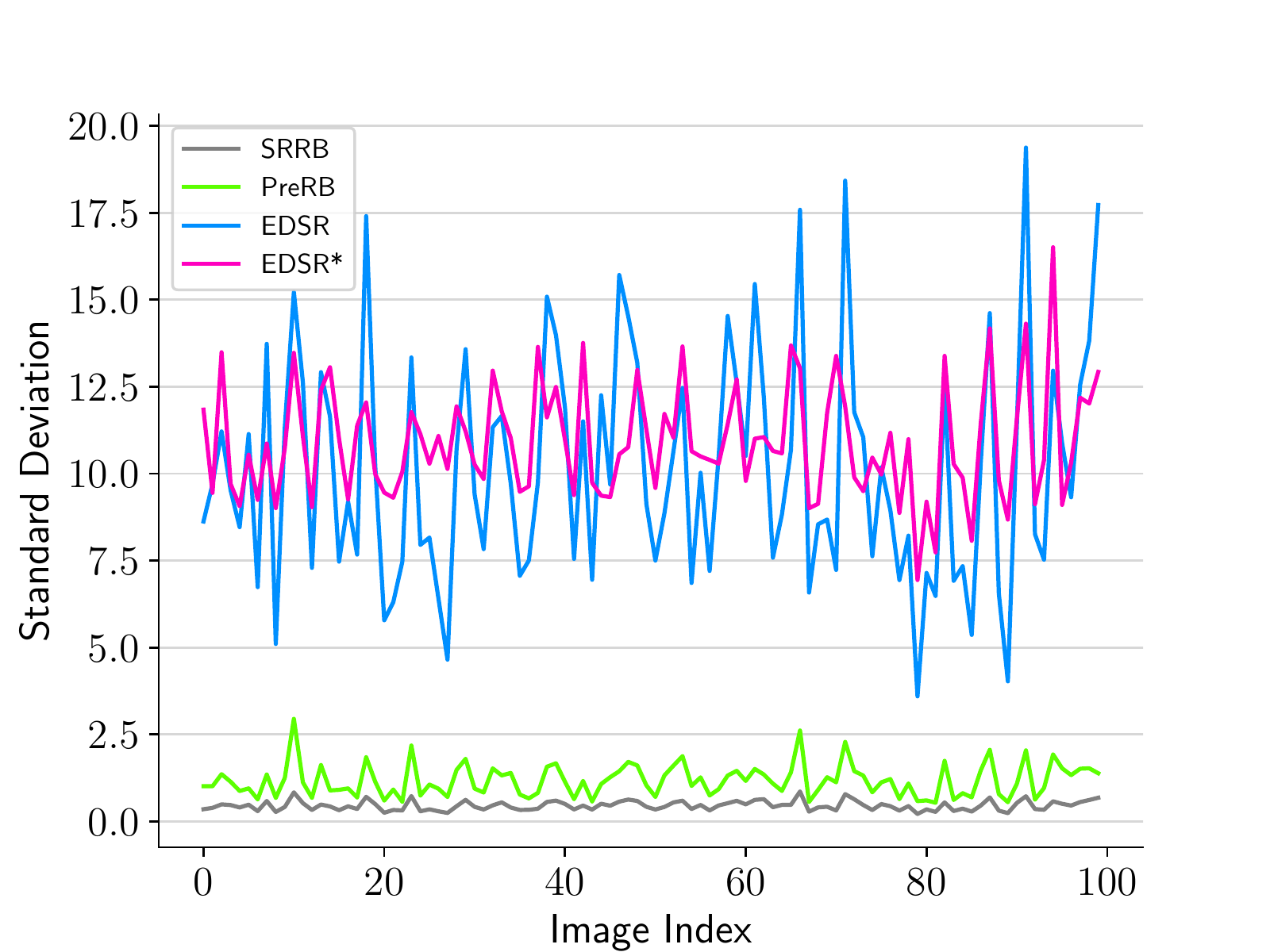}
        \caption{}\label{fig:std_curve}
    \end{subfigure}
    \begin{subfigure}[b]{0.45\linewidth}
        \centering
        \includegraphics[width=\linewidth]{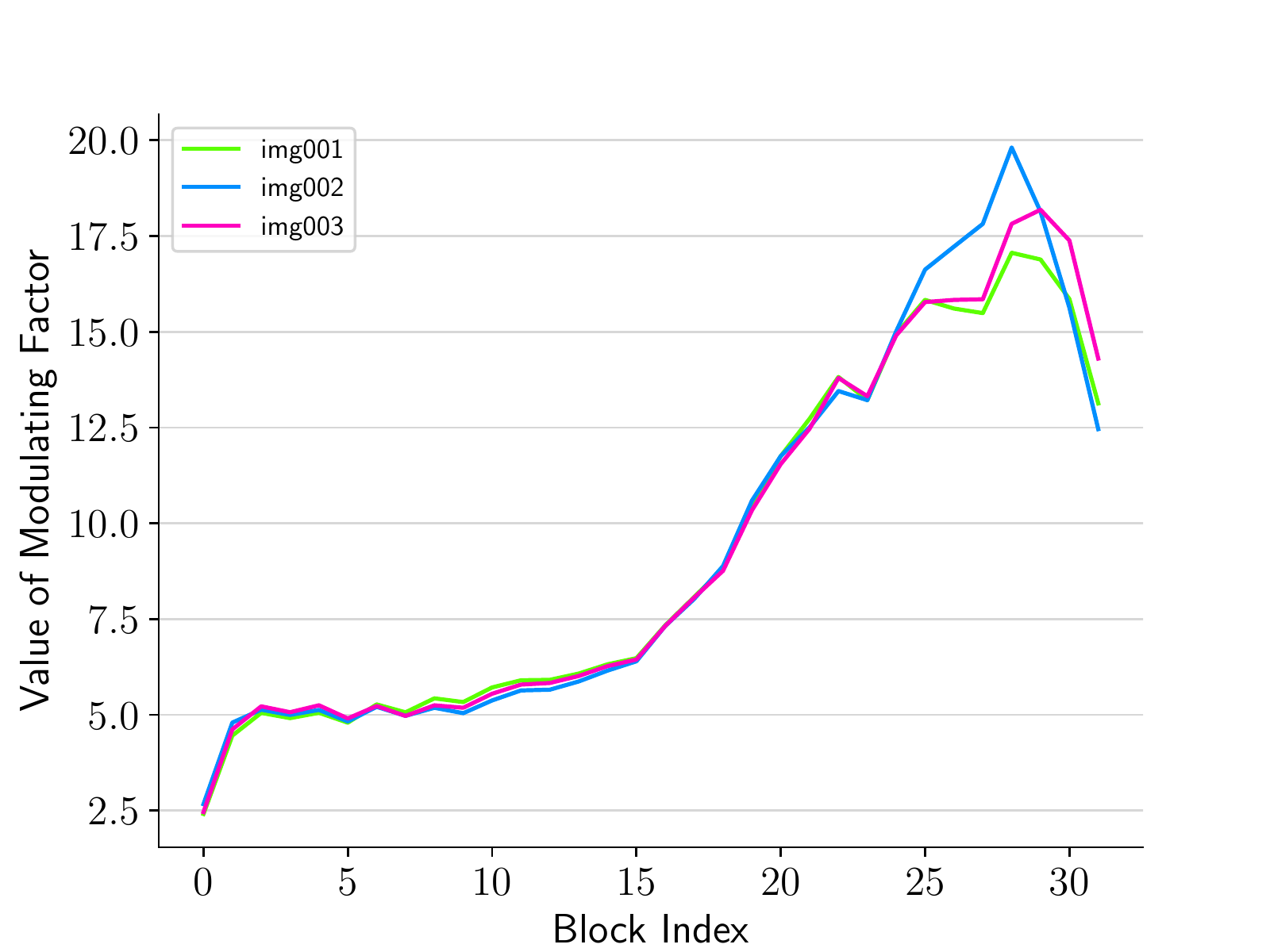}
        \caption{}\label{fig:AdaDM_scale}
    \end{subfigure}
    \caption{(a) Deviation comparisons of different models on the 100 images from Urban100 $\times 2$ dataset.
        The standard deviation is computed on the residual feature (\ie, the feature before element-wise addition) of 
        the last residual block. (b) Modulation factor in each residual block of EDSR* for the first three images from
        Urban100 $\times 2$ dataset.}
\end{figure*}

\section{Experiments}
\subsection{Datasets and Metrics}
Following~\cite{EDSR,RDN,NLSN}, 800 training images from DIV2K~\cite{DIV2K}
are used as the training dataset. All the models are evaluated on five benchmark datasets: Set5~\cite{Set5},
Set14~\cite{Set14}, B100~\cite{B100}, Urban100~\cite{Urban100} and Manga109~\cite{Manga109}. The SR results
are evaluated with PSNR and SSIM~\cite{SSIM} on Y channel of transformed YCbCr space.

\subsection{Training and Implementation Details}
As with previous works~\cite{EDSR,RDN,NLSN}, data augmentation is performed on the 800 training images, which are 
randomly rotated by $90^\circ$, $180^\circ$, $270^\circ$ and flipped horizontally. In each mini-batch, 16 LR
color patches with size $48\times 48$ are cropped as the inputs. All the models are trained by ADAM~\cite{ADAM}
optimizer with $\beta_1 = 0.9$, $\beta_2 = 0.999$, and $\epsilon = 10^{-8}$. The learning rate is set to $10^{-4}$
and reduced to half every 200 epochs. The final model is obtained after 1000 epochs. We use the PyTorch
framework to train our models with a Tesla V100 GPU. For our AdaDM, the weight $w$ is initialized to 1 and the 
bias $b$ is initialized to 0 as discussed in \S~\ref{sec:AdaDM}.

\subsection{Ablation Study}
In this section, we conduct experiments to study the impact of individual components. For comparison, we train the
EDSR model as a baseline, which is included in the second column of Table~\ref{tab:RB_SRRB}, Table~\ref{tab:RB_SRRBPre}
and Table~\ref{tab:ablation}. 

\noindent\textbf{Positions of BN.} In \cite{EDSR}, the authors have discussed EDSR with and without BN layers.
EDSR with BN actually uses the residual block in SRResNet~\cite{SRResNet}, where a BN layer is added
after each convolution. We refer to the residual block in EDSR as ``RB'' (Figure~\ref{fig:RB}) and the residual block in
SRResNet as ``SRRB'' (Figure~\ref{fig:SRResNet}).
Different from SRRB, we add the BN layers in front of the convolutional layers in collaboration with our AdaDM. 
We refer to this residual block as ``PreRB'', which is depicted in Figure~\ref{fig:AdaDM_BN}. 
It can be observed from Table~\ref{tab:RB_SRRB} and Table~\ref{tab:RB_SRRBPre} that SRRB performs much worse than RB while
PreRB achieves similar results with RB. It indicates that we should place the BN layer before the convolutional layer 
in SR network. 
In PreRB, each convolution can benefit from normalized input feature and the last convolution can adjust 
the pixel deviation to a certain extent before adding to the identity branch. As a consequence, it performs better than SRRB.
Though PreRB can keep the performance, it does not bring much improvement on benchmark datasets because of the reduced deviation.
Our AdaDM can address this problem and show the usefulness of BN to improve SR performance.

\noindent\textbf{Effects of AdaDM.} We investigate the effects of AdaDM by removing all BN layers
in EDSR*, \ie, ``AdaDM only'' in Table~\ref{tab:ablation}. 
In experiments, we found that AdaDM only does not converge at all because of huge loss values. Therefore, our AdaDM cannot work alone as a residual 
scaling strategy. On the contrary, it can work collaboratively with normalization layers, which demonstrates that the contribution of
our AdaDM comes from deviation amplification rather than residual scaling. The last column of Table~\ref{tab:ablation} shows the results of
our full model. It outperforms the baseline EDSR model by a large margin on Urban100 (\textbf{+0.16}dB) and Manga109 (\textbf{+0.18}dB)
datasets, which further proves the effectiveness of our AdaDM.

\subsection{Analysis of Deviation Amplification}
In this part, we give more analysis of the deviation amplification mechanism in AdaDM. Figure~\ref{fig:std_curve} depicts the deviation distributions
for the models in ablation study. 
By comparing the four curves, we can conclude two factors that affect the performance of SR networks:

\noindent\textbf{Amplitude of deviation}. As shown in the bottom of Figure~\ref{fig:std_curve}, PreRB has more amplitude changes
than SRRB. Besides, we have also known from Table~\ref{tab:RB_SRRB} and Table~\ref{tab:RB_SRRBPre} that PreRB has much better performance
than SRRB. Based on these two observations, we can infer that the deviation amplitude plays a critical role in achieving good SR performance.
Similar observations can also be obtained by comparing EDSR* and PreRB. EDSR* enhances the amplitude of PreRB via the proposed AdaDM.

\begin{figure}
    \centering
    \includegraphics[width=0.85\linewidth]{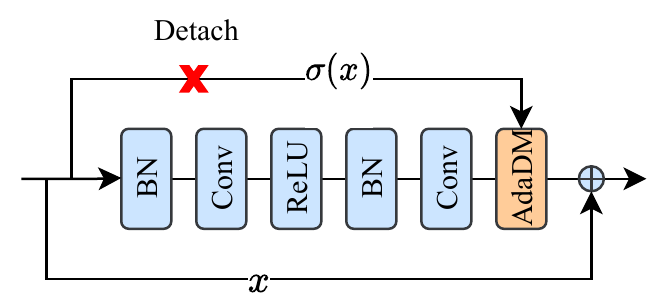}
    \caption{Detached AdaDM. The backpropagated gradients via the upper skip connection are detached.}\label{fig:AdaDM_detach}
\end{figure}

\begin{table}[t]
    \centering
    \caption{Quantitative comparison (average PSNR / SSIM) of AdaDM and detached AdaDM on benchmark datasets
    for $\times 2$ SR.}\label{tab:AdaDM_detach}
    \resizebox{\linewidth}{!}{
    \begin{tabular}{lccccc}
    \toprule
    Method & Set5 &  Set14 & B100 & Urban100 & Manga109 \\
    \midrule
    AdaDM & 38.25 / {\color{red}0.9615} & 34.00 / 0.9205 & 32.37 / 0.9022 & {\color{red}33.12} / {\color{red}0.9371} &  {\color{red}39.31} / {\color{red}0.9783} \\
    Detached & {\color{red}38.26} / 0.9614 & {\color{red}34.10} / {\color{red}0.9215} & {\color{red}32.38} / {\color{red}0.9023} & 33.10 / {\color{red}0.9371} & 39.26 / 0.9780 \\
    \bottomrule
    \end{tabular}}


    \caption{Quantitative comparison (average PSNR / SSIM) of CRAN, DFSA and NLSN* with different training datasets
    for $\times 2$ SR. ``$\dag$'' means the results are take from their publications.}\label{tab:df2k}
    \resizebox{\linewidth}{!}{
    \begin{tabular}{lcccccc}
    \toprule
    Method & Dataset & Set5 &  Set14 & B100 & Urban100 & Manga109 \\
    \midrule
    NLSN\dag~\cite{NLSN}  & DIV2K& 38.34 / 0.9618 & 34.08 / 0.9231 & 32.43 / 0.9027 & 33.42 / 0.9394 & 39.59 / 0.9789\\
    NLSN* & DIV2K  & {\color{blue}38.39} / 0.9619 & 34.21 / {\color{blue}0.9236} & {\color{blue}32.46} / {\color{blue}0.9031} & 33.59 / 0.9410 & 39.67 / 0.9791 \\
    \midrule
    CRAN\dag~\cite{CRAN} & DF2K & 38.31 / 0.9617 & 34.22 / 0.9232 & 32.44 / 0.9029 & 33.43 / 0.9394 & 39.75 / {\color{blue}0.9793} \\
    DFSA\dag~\cite{DFSA} & DF2K & 38.38 / {\color{blue}0.9620} & {\color{blue}34.33} / 0.9232 & {\color{red}32.50} / {\color{red}0.9036} & {\color{blue}33.66} / {\color{blue}0.9412} & {\color{red}39.98} / {\color{red}0.9798} \\
    \hdashline
    NLSN* & DF2K  & {\color{red}38.43} / {\color{red}0.9622} & {\color{red}34.40} / {\color{red}0.9249} & {\color{red}32.50} / {\color{red}0.9036} & {\color{red}33.78} / {\color{red}0.9419} & {\color{blue}39.89} / {\color{red}0.9798} \\
    \bottomrule
    \end{tabular}}

    \caption{Complexity comparisons of EDSR and EDSR* on Urban100 dataset
        for $\times 2$ SR. Test time is estimated with the ``torch.cuda.Event()'' function and GPU memory is estimated 
    with the ``torch.cuda.max\_memory\_allocated()'' function.}\label{tab:time_memory}
    \resizebox{\linewidth}{!}{
    \begin{tabular}{lcccccc}
    \toprule
    Method & Params &  Mult-Adds & Test Time & Test Memory & Training Memory & PSNR\\
    \midrule
    EDSR & 40.73M & 2.35T & 0.72s/img & 3342.74M &  3813.74M & 32.96dB \\
    EDSR* & 40.76M & 2.35T & 0.75s/img & 3343.07M & 7437.57M & 33.12dB\\
    \bottomrule
    \end{tabular}}

\end{table}

\noindent\textbf{Steadiness of deviation}. EDSR* has much better performance than EDSR while its amplitude changes are smaller than the latter,
which seems to violate the first conclusion. Here, we think that the performance improvements come from the regularization effects of 
normalization layers. Because of BN, EDSR* has a more steady deviation distribution across different images, which shows the necessity of using
normalization in SR networks. This observation can also be used to explain why PreRB has a similar performance with EDSR
in Table~\ref{tab:RB_SRRBPre}. Though PreRB has smaller deviation amplitudes, it has more steady deviation changes across test images.
In other words, SR networks with normalization layers can have better generalization ability.

In Figure~\ref{fig:AdaDM_scale}, we further visualize the modulation factors in each residual block of EDSR*. For different images, the network
tries to conduct similar deviation amplification at the beginning and finally predicts totally different factors for each test image, which indicates
that EDSR* first exploits common feature representations and then learns specific features for HR image reconstruction.
\begin{figure*}
    \centering
    \begin{subfigure}[b]{0.48\linewidth}
        \centering
        \includegraphics[width=\linewidth]{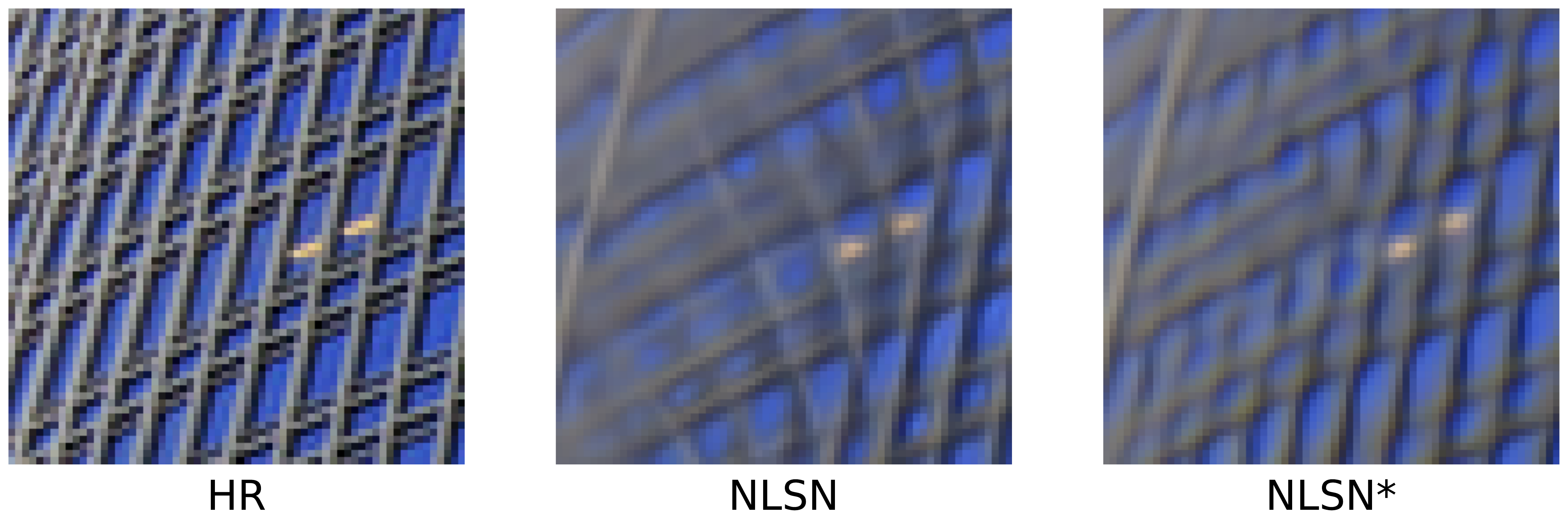}
    \end{subfigure}
    \begin{subfigure}[b]{0.48\linewidth}
        \centering
        \includegraphics[width=\linewidth]{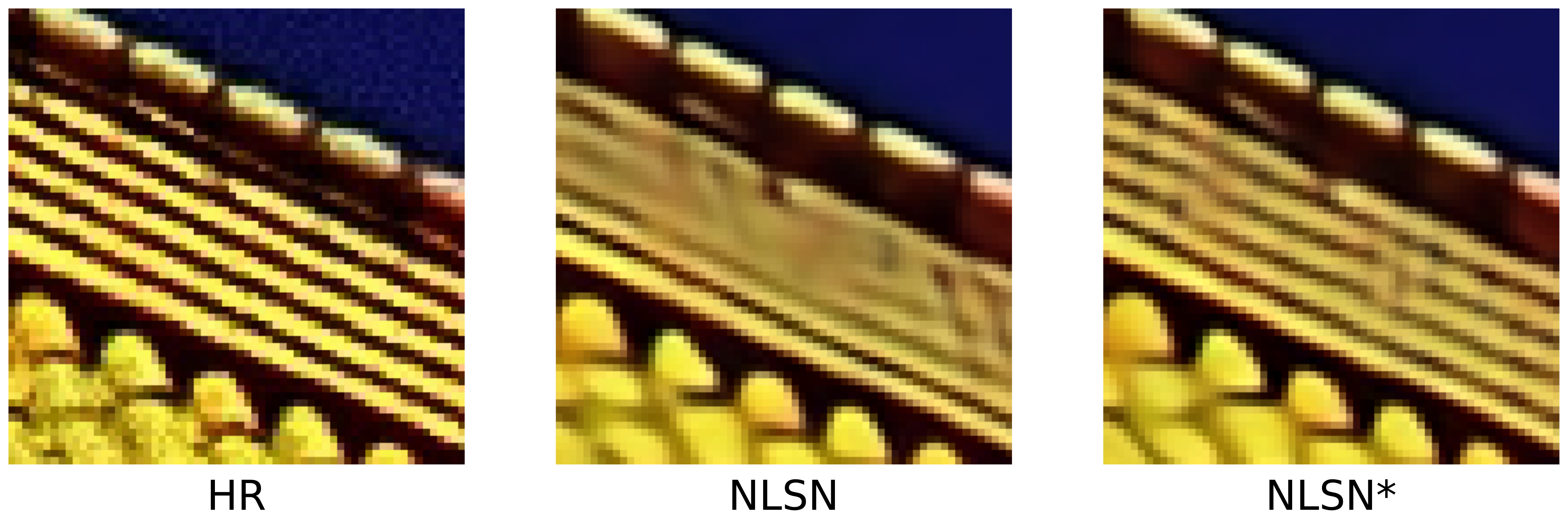}
    \end{subfigure}
    \caption{Visual comparisons of NLSN and NLSN* on Urban100 dataset for $\times 4$ SR.}\label{fig:demo}
\end{figure*}
\begin{table*}[htp]
    \centering
    \caption{Quantitative comparison (average PSNR / SSIM) with state-ot-the-art methods on benchmark datasets. ``$\dag$'' means the results 
    are take from their publications, otherwise the results are from models retrained under the same experimental settings.}
    \label{tab:sota}
    \resizebox{0.95\textwidth}{!}{
    \begin{tabular}{|l|c|c|c|c|c|c|c|}
        \hline
        \multirow{2}{*}{Method} & \multirow{2}{*}{Scale} & \multirow{2}{*}{Training Dataset} & Set5 & Set14 & BSD100 & Urban100 & Manga109\\
        \cline{4-8}
         & & & PSNR/SSIM & PSNR/SSIM & PSNR/SSIM & PSNR/SSIM & PSNR/SSIM\\
        \hline \hline
        RCAN\dag~\cite{RCAN} & $\times 2$ &DIV2K & 38.27 / 0.9614 & 34.12 / 0.9216 & 32.41 / 0.9027 & 33.34 / 0.9384 & 39.44 / 0.9786 \\
        SAN\dag~\cite{SAN} & $\times 2$ & DIV2K & 38.31 / {\color{red}0.9620} & 34.07 / 0.9213 & 32.42 / {\color{blue}0.9028} & 33.10 / 0.9370 & 39.32 / {\color{red}0.9792}\\
        RFANet\dag~\cite{RFANet} & $\times 2$ & DIV2K & 38.26 / 0.9615 & {\color{blue}34.16} / 0.9220 & 32.41 / 0.9026 & 33.33 / 0.9389 & 39.44 / 0.9783\\
        HAN\dag~\cite{HAN} & $\times 2$ & DIV2K & 38.27 / 0.9614 & {\color{blue}34.16} / 0.9217 & 32.41 / 0.9027 & 33.35 / 0.9385 & 39.46 / 0.9785\\
        IGNN\dag~\cite{IGNN} & $\times 2$ & DIV2K & 38.24 / 0.9613 & 34.07 / 0.9217 & 32.41 / 0.9025 & 33.23 / 0.9383 & 39.35 / 0.9786\\
        EDSR\dag~\cite{EDSR} & $\times 2$ & DIV2K& 38.11 / 0.9602 & 33.92 / 0.9195 & 32.32 / 0.9013 & 32.93 / 0.9351 & 39.10 / 0.9773\\
        RDN\dag~\cite{RDN} & $\times 2$ &DIV2K & 38.24 / 0.9614 & 34.01 / 0.9212 & 32.34 / 0.9017 & 32.89 / 0.9353 & 39.18 / 0.9780\\
        NLSN\dag~\cite{NLSN} & $\times 2$ & DIV2K& {\color{blue}38.34} / {0.9618} & 34.08 / {\color{blue}0.9231} & {\color{blue}32.43} / 0.9027 & 33.42 / 0.9394 & {\color{blue}39.59} / {0.9789}\\
        \hline
        EDSR & $\times 2$ &DIV2K & 38.22 / 0.9612 & 33.97 / 0.9207 & 32.35 / 0.9019 & 32.96 / 0.9359 & 39.13 / 0.9778\\
        EDSR* & $\times 2$ &DIV2K & 38.25 / 0.9615 & 34.00 / 0.9205 & 32.37 / 0.9022 & 33.12 / 0.9371 & 39.31 / 0.9783\\
        \hdashline
        RDN & $\times 2$ & DIV2K& 38.22 / 0.9613 & 33.93 / 0.9207 & 32.34 / 0.9017 & 32.89 / 0.9352 & 39.13 / 0.9780\\
        RDN* & $\times 2$ &DIV2K & 38.22 / 0.9612 & 33.99 / 0.9214 & 32.36 / 0.9019 & 33.03 / 0.9365 & 39.18 / 0.9778\\
        \hdashline
        NLSN & $\times 2$ &DIV2K & 38.33 / {0.9618} & {\color{red}34.21} / 0.9228 & {\color{blue}32.43} / {\color{blue}0.9028} & {\color{blue}33.45} / {\color{blue}0.9396} & 39.53 / 0.9787\\
        NLSN* & $\times 2$ &DIV2K & {\color{red}38.39} / {\color{blue}0.9619} & {\color{red}34.21} / {\color{red}0.9236} & {\color{red}32.46} / {\color{red}0.9031} & {\color{red}33.59} / {\color{red}0.9410} & {\color{red}39.67} / {\color{blue}0.9791}\\
        \hline \hline
        
        RCAN\dag~\cite{RCAN} & $\times 3$ &DIV2K & 34.74 / 0.9299 & 30.65 / 0.8482 & 29.32 / 0.8111 & 29.09 / 0.8702 & 34.44 / 0.9499 \\
        SAN\dag~\cite{SAN} & $\times 3$ &DIV2K & 34.75 / 0.9300 & 30.59 / 0.8476 & 29.33 / 0.8112 & 28.93 / 0.8671 & 34.30 / 0.9494\\
        RFANet\dag~\cite{RFANet} & $\times 3$ & DIV2K& 34.79 / 0.9300 & 30.67 / 0.8487 & 29.34 / 0.8115 & 29.15 / 0.8720 & 34.59 / 0.9506\\
        HAN\dag~\cite{HAN} & $\times 3$ &DIV2K & 34.75 / 0.9299 & 30.67 / 0.8483 & 29.32 / 0.8110 & 29.10 / 0.8705 & 34.48 / 0.9500\\
        IGNN\dag~\cite{IGNN} & $\times 3$ &DIV2K & 34.72 / 0.9298 & 30.66 / 0.8484 & 29.31 / 0.8105 & 29.03 / 0.8696 & 34.39 / 0.9496\\
        EDSR\dag~\cite{EDSR} & $\times 3$ &DIV2K & 34.65 / 0.9280 & 30.52 / 0.8462 & 29.25 / 0.8093 & 28.80 / 0.8653 & 34.17 / 0.9476\\
        RDN\dag~\cite{RDN} & $\times 3$ &DIV2K & 34.71 / 0.9296 & 30.57 / 0.8468 & 29.26 / 0.8093 & 28.80 / 0.8653 & 34.13 / 0.9484\\
        NLSN\dag~\cite{NLSN} & $\times 3$ &DIV2K & 34.85 / 0.9306 & 30.70 / 0.8485 & 29.34 / 0.8117 & 29.25 / 0.8726 & 34.57 / 0.9508\\
        \hline
        EDSR & $\times 3$ &DIV2K & 34.70 / 0.9295 & 30.56 / 0.8465 & 29.26 / 0.8097 & 28.85 / 0.8663 & 34.05 / 0.9483\\
        EDSR* & $\times 3$ &DIV2K & 34.81 / 0.9302 & 30.63 / 0.8481 & 29.31 / 0.8108 & 29.02 / 0.8693 & 34.48 / 0.9499\\
        \hdashline
        RDN & $\times 3$ &DIV2K & 34.70 / 0.9295 & 30.56 / 0.8464 & 29.26 / 0.8092 & 28.80 / 0.8654 & 34.11 / 0.9483\\
        RDN* & $\times 3$ & DIV2K& 34.79 / 0.9300 & 30.62 / 0.8477 & 29.28 / 0.8097 & 28.95 / 0.8676 & 34.29 / 0.9490\\
        \hdashline
        NLSN & $\times 3$ & DIV2K & {\color{blue}34.90} / {\color{blue}0.9310} & {\color{blue}30.74} / {\color{blue}0.8495} & {\color{blue}29.37} / {\color{blue}0.8126} & {\color{blue}29.34} / {\color{blue}0.8746} & {\color{blue}34.64} / {\color{blue}0.9513}\\
        NLSN* & $\times 3$ & DIV2K & {\color{red}34.94} / {\color{red}0.9313} & {\color{red}30.80} / {\color{red}0.8503} & {\color{red}29.40} / {\color{red}0.8130} & {\color{red}29.53} / {\color{red}0.8775} & {\color{red}34.95} / {\color{red}0.9524}\\
        \hline \hline
        RCAN\dag~\cite{RCAN} & $\times 4$ &DIV2K & 32.63 / 0.9002 & 28.87 / 0.7889 & 27.77 / 0.7436 & 26.82 / 0.8087 & 31.22 / 0.9173 \\
        SAN\dag~\cite{SAN} & $\times 4$ &DIV2K & 32.64 / 0.9003 & 28.92 / 0.7888 & 27.78 / 0.7436 & 26.79 / 0.8068 & 31.18 / 0.9169\\
        RFANet\dag~\cite{RFANet} & $\times 4$ &DIV2K & {\color{blue}32.66} / 0.9004 & 28.88 / 0.7894 & 27.79 / 0.7442 & 26.92 / 0.8112 & 31.41 / 0.9187\\
        HAN\dag~\cite{HAN} & $\times 4$ &DIV2K & 32.64 / 0.9002 & 28.90 / 0.7890 & 27.80 / 0.7442 & 26.85 / 0.8094 & 31.42 / 0.9177\\
        IGNN\dag~\cite{IGNN} & $\times 4$ &DIV2K & 32.57 / 0.8998 & 28.85 / 0.7891 & 27.77 / 0.7434 & 26.84 / 0.8090 & 31.28 / 0.9182\\
        EDSR\dag~\cite{EDSR} & $\times 4$ &DIV2K & 32.46 / 0.8968 & 28.80 / 0.7876 & 27.71 / 0.7420 & 26.64 / 0.8033 & 31.02 / 0.9148\\
        RDN\dag~\cite{RDN} & $\times 4$ &DIV2K & 32.47 / 0.8990 & 28.81 / 0.7871 & 27.72 / 0.7419 & 26.61 / 0.8028 & 31.00 / 0.9151\\
        NLSN$\dag$~\cite{NLSN} & $\times 4$ & DIV2K& 32.59 / 0.9000 & 28.87 / 0.7891 & 27.78 / 0.7444 & 26.96 / 0.8109 & 31.27 / 0.9184\\
        \hline
        EDSR & $\times 4$ &DIV2K & 32.51 / 0.8985 & 28.78 / 0.7872 & 27.73 / 0.7419 & 26.66 / 0.8040 & 31.04 / 0.9152\\
        EDSR* & $\times 4$ &DIV2K & 32.59 / 0.8995 & 28.87 / 0.7887 & 27.76 / 0.7433 & 26.83 / 0.8079 & 31.24 / 0.9172\\
        \hdashline
        RDN & $\times 4$ &DIV2K & 32.48 / 0.8988 & 28.81 / 0.7869 & 27.72 / 0.7411 & 26.62 / 0.8023 & 31.03 / 0.9154\\
        RDN* & $\times 4$ &DIV2K & 32.49 / 0.8991 & 28.84 / 0.7883 & 27.74 / 0.7423 & 26.72 / 0.8056 & 31.18 / 0.9171\\
        \hdashline
        NLSN & $\times 4$ &DIV2K & {32.65} / {\color{blue}0.9009} & {\color{blue}28.93} / {\color{blue}0.7900} & {\color{blue}27.82} / {\color{blue}0.7453} & {\color{blue}27.06} / {\color{blue}0.8147} & {\color{blue}31.43} / {\color{blue}0.9202}\\
        NLSN* & $\times 4$ &DIV2K & {\color{red}32.75} / {\color{red}0.9018} & {\color{red}28.96} / {\color{red}0.7917} & {\color{red}27.85} / {\color{red}0.7464} & {\color{red}27.24} / {\color{red}0.8182} & {\color{red}31.73} / {\color{red}0.9225}\\
        \hline 
\end{tabular}}
\end{table*}


\subsection{Impact of Skip Connection}
As shown in Figure~\ref{fig:RB_AdaDM}, we add a skip connection to compute the standard deviation of input feature. The standard deviation is used
in our AdaDM for modulation factor regression. Though our AdaDM achieves substantial improvements, it is unclear whether the improvements come
from deviation amplification or the added skip connection that affects the gradient backpropagation. The upper skip connection and the computation
of standard deviation will bring extra gradients to the input.
To eliminate this concern, we train a comparison EDSR model
that uses the detached AdaDM in Figure~\ref{fig:AdaDM_detach}. We detach the gradients of the upper skip connection during backpropagation.
The comparison results are included in Table~\ref{tab:AdaDM_detach}. We can observe that the two models perform similarly on Set5, B100 and Urban100 
datasets. The detached AdaDM is 0.1dB higher on Set14 dataset and is 0.05dB lower on Manga109 dataset, which means that the detached AdaDM 
even has slightly better performance. Based on these observations, we can conclude that the main contribution comes from our deviation amplification during forward propagation.  

\subsection{Comparisons with State-of-the-Arts}
The comparisons with state-of-the-art image SR models are shown in Table~\ref{tab:sota}. 
In Table~\ref{tab:sota}, EDSR*, RDN* and NLSN* denote the model using
BN layers and our AdaDM in their building blocks, \ie, the blocks depicted in Figure~\ref{fig:RB_AdaDM} and Figure~\ref{fig:RDB_AdaDM}.
For EDSR, RDN and NLSN, we include both the results from their publications (mark with ``$\dag$'') and the results reproduced by us.
As we can see, EDSR*, RDN* and NLSN* have much higher PSNR than their counterparts. For example, NLSN* achieves 0.3dB improvements on 
Manga109 dataset for $\times 4$ SR. 
We also show the visual comparisons of NLSN and NLSN* in Figure~\ref{fig:demo}. The NLSN* can reconstruct more edge details than original NLSN.


\subsection{About the Training Dataset}
In this paper, we only use 800 training images from DIV2K as the training dataset and no extra training dataset
is used. To show this, we train the $\text{NLSN}^*$ model with ``DIV2K+Flickr2K'' (DF2K) dataset for $\times 2$ SR.
The evaluation results are shown in Table~\ref{tab:df2k}. NLSN* trained on DIV2K and Flickr2K~\cite{EDSR} datasets shows much better performance than
the model trained on DIV2K dataset. Moreover, NLSN* has much better performance than CRAN~\cite{CRAN} and outperforms DFSA~\cite{DFSA} on most cases.

\section{Complexity Analysis}
Table~\ref{tab:time_memory} shows the complexity comparisons of EDSR and EDSR*. As we can see, EDSR and EDSR* have similar test time and 
test memory consumption. The main limitation is the training memory required by EDSR*. If the training time memory consumption is not a problem,
the proposed normalization technique can bring a significant performance improvement while not affecting the test time and memory much.
\begin{figure*}[t]
    \centering
    \begin{subfigure}[b]{0.495\linewidth}
        \centering
    \includegraphics[width=\linewidth]{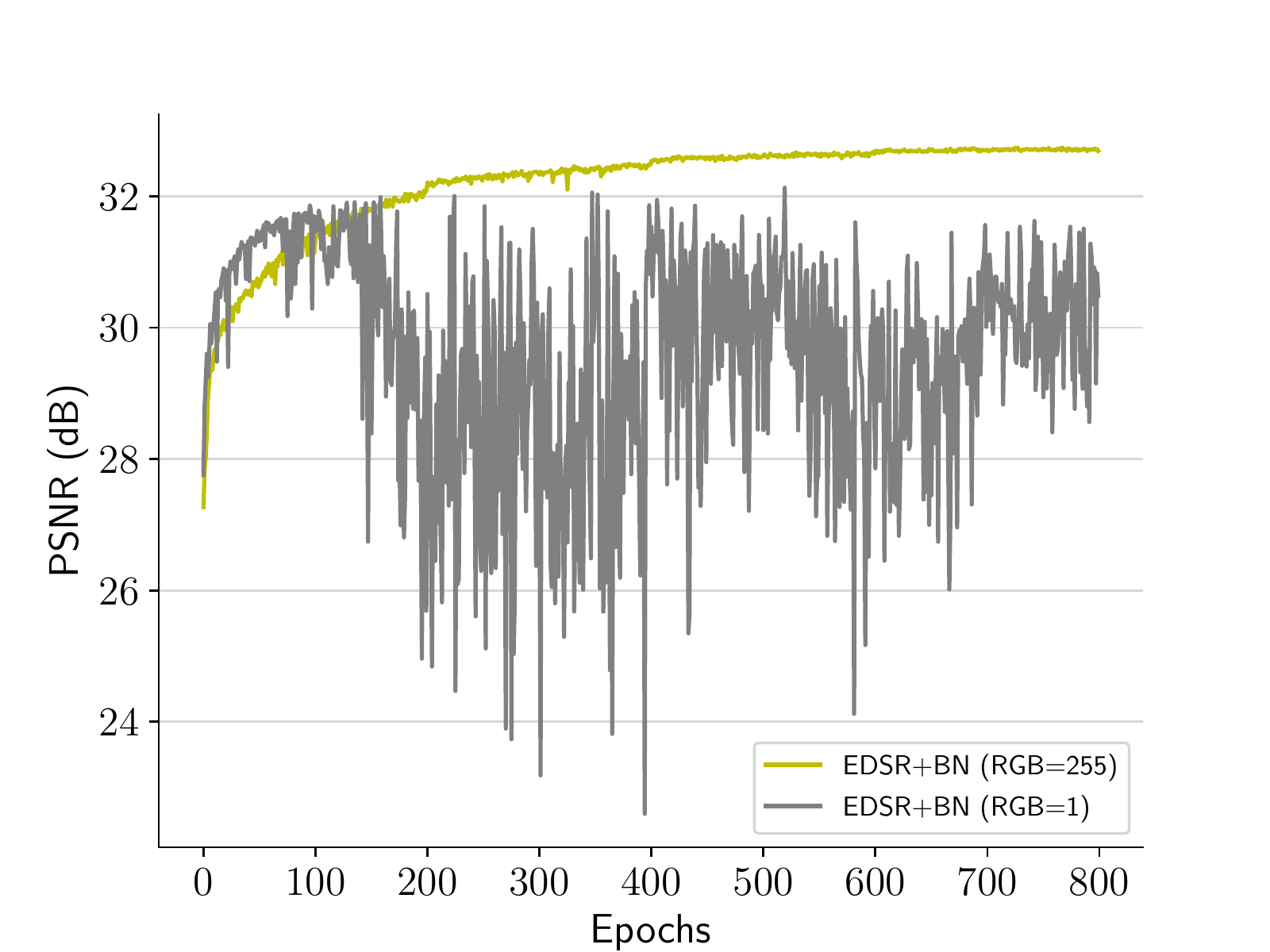}
    \caption{}\label{fig:RB}\label{fig:rgb_bn}
    \end{subfigure}
    \begin{subfigure}[b]{0.495\linewidth}
        \centering
    \includegraphics[width=\linewidth]{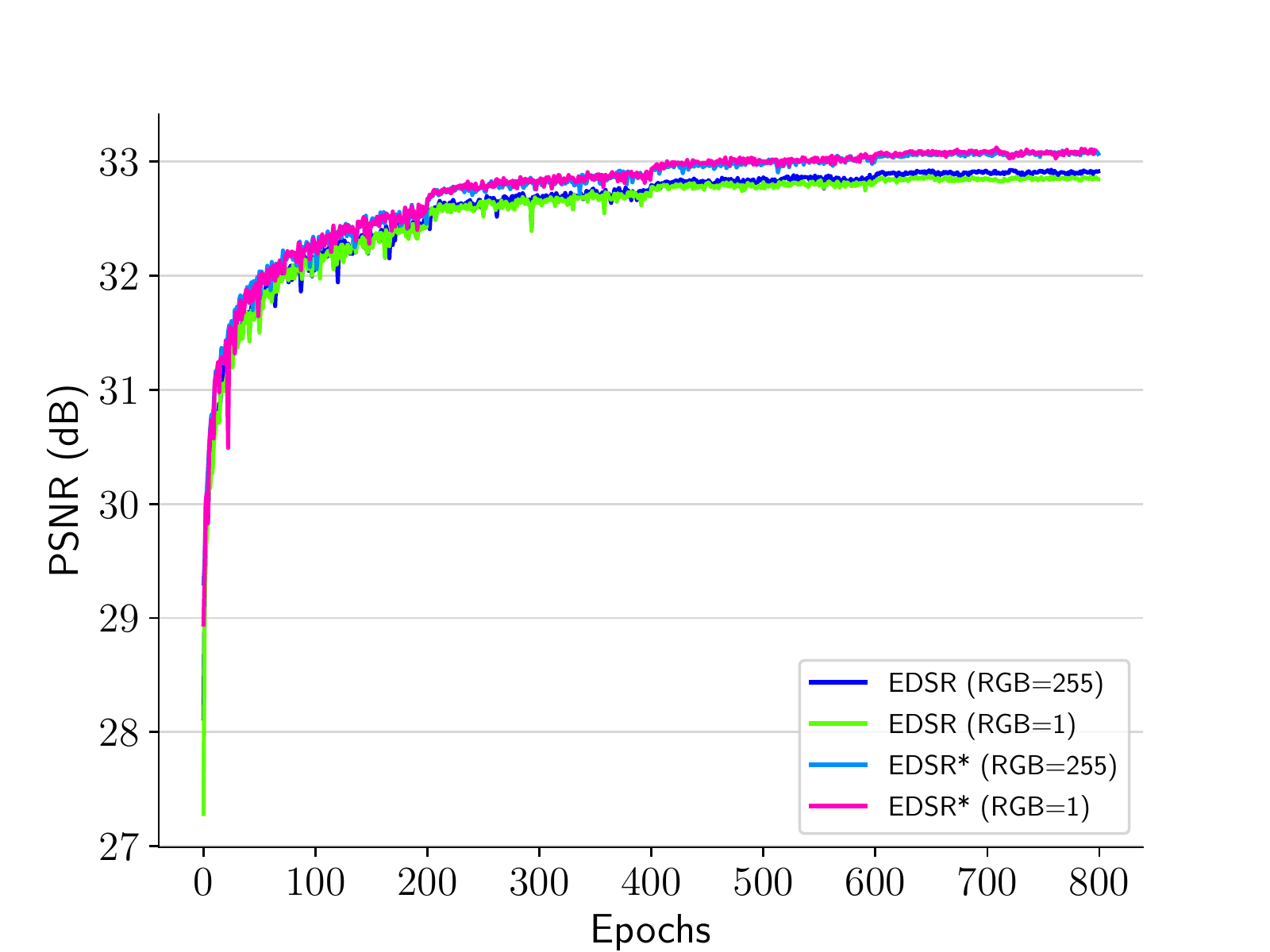}
    \caption{}\label{fig:RB_AdaDM}\label{fig:rgb_range}
    \end{subfigure}
    \caption{Training curves of EDSR, EDSR+BN and EDSR*.}
\end{figure*}

\section{Conclusion}
In this paper, we give a comprehensive analysis of feature normalization in image Super-Resolution (SR) networks. We found that the performance
degradation caused by normalization layers comes from the reduced pixel deviation of residual features. When the pixel deviation shrinks, 
the variation of pixel values becomes smaller, making the edges less discriminative for the network to resolve. To address this problem,
we propose an Adaptive Deviation Modulator (AdaDM) that can amplify the pixel deviation of normalized residual features. Thanks to our 
AdaDM, we successfully apply normalization layers to state-of-the-art SR networks and achieve substantial performance improvements.

\section*{A1. About the RGB Range}
As we have analyzed in the paper that normalization layers in SR networks will reduce the pixel deviation by dividing by a $\sigma$. 
In practice, we can adopt the RGB range of 255 (\ie, $[0,255]$) or 1 (\ie, $[0,1]$) for model training. When we choose the training RGB range of 1, pixel
scaling occurs implicitly. In the following, we will investigate whether the training RGB range has an impact on performance.
Figure~\ref{fig:rgb_bn} shows the training curves of EDSR+BN with RGB range of 255 and 1, respectively. When training with RGB range of 1,
EDSR+BN becomes unstable and it converges much worse than training with RGB range of 255. Therefore, it is better to use RGB range of 255 
when training with BN in SR networks since BN will further reduce the pixel deviation. Figure~\ref{fig:rgb_range} shows the training curves of original EDSR and EDSR*. Different from EDSR+BN, the training RGB range has little impact on original EDSR model since it removes the BN layers in residual blocks.
Our EDSR* has almost the same performance when training with different RGB ranges, which indicates that our AdaDM is effective enough in 
deviation amplification and it is robust to training RGB ranges.

\section*{A2. Results for DF2K Dataset}
\begin{table}[h]
    \vspace{-1.0em}
    \centering
    \caption{Quantitative comparison (average PSNR / SSIM) of NLSN~\cite{NLSN} and SwinIR~\cite{SwinIR} with different training datasets
    for $\times 2$ SR. ``+'' means the results are obtained with self-ensemble testing.}\label{tab:df2k_x2}
    \resizebox{\linewidth}{!}{
    \begin{tabular}{lcccccc}
    \toprule
    Method & Dataset & Set5 &  Set14 & B100 & Urban100 & Manga109 \\
    \midrule
    NLSN* & DIV2K  & {38.39} / 0.9619 & 34.21 / {0.9236} & {32.46} / {0.9031} & 33.59 / 0.9410 & 39.67 / 0.9791 \\
    NLSN*+  & DIV2K& 38.43 / 0.9621 & 34.37 / 0.9245 & 32.50 / 0.9036 & 33.83 / 0.9424 & 39.83 / 0.9795\\
    \midrule
    SwinIR+ & DF2K & {\color{blue}38.46} / {\color{red}0.9624} & {\color{red}34.61} / {\color{red}0.9260} & {\color{red}32.55} / {\color{red}0.9043} & {\color{blue}33.95} / {\color{red}0.9433} & {\color{blue}40.02} / {\color{blue}0.9800} \\
    \hdashline
    NLSN* & DF2K  & {38.43} / {\color{blue}0.9622} & {34.40} / {0.9249} & {32.50} / {0.9036} & {33.78} / {0.9419} & {39.89} / {0.9798} \\
    NLSN*+ & DF2K  & {\color{red}38.50} / {\color{red}0.9624} & {\color{blue}34.46} / {\color{blue}0.9255} & {\color{blue}32.53} / {\color{blue}0.9040} & {\color{red}33.96} / {\color{blue}0.9431} & {\color{red}40.03} / {\color{red}0.9801} \\
    \bottomrule
    \end{tabular}}
    \quad
    \caption{Quantitative comparison (average PSNR / SSIM) of NLSN~\cite{NLSN} and SwinIR~\cite{SwinIR} with different training datasets
    for $\times 3$ SR. ``+'' means the results are obtained with self-ensemble testing.}\label{tab:df2k_x3}
    \resizebox{\linewidth}{!}{
    \begin{tabular}{lcccccc}
    \toprule
    Method & Dataset & Set5 &  Set14 & B100 & Urban100 & Manga109 \\
    \midrule
    NLSN* & DIV2K  & 34.94 / 0.9313 & 30.80 / 0.8503 & 29.40 / 0.8130 & 29.53 / 0.8775 & 34.95 / 0.9524 \\
    NLSN*+  & DIV2K& {\color{blue}35.00} /0.9318 & 30.89 / 0.8519 & {\color{blue}29.45} / 0.8139 & 29.74 / 0.8803 & 35.19 / 0.9535\\
    \midrule
    SwinIR+ & DF2K & {\color{red}35.04} / {\color{red}0.9322} & {\color{red}31.00} / {\color{red}0.8542} & {\color{red}29.49} / {\color{red}0.8150} & {\color{blue}29.90} / {\color{red}0.8841} & {\color{blue}35.28} / {\color{red}0.9543} \\
    \hdashline
    NLSN* & DF2K & 34.95 / 0.9316 & 30.86 / 0.8513 & {\color{blue}29.45} / 0.8141 & 29.77 / {0.8812} & 35.20 / 0.9534 \\
    NLSN*+ & DF2K  & {\color{red}35.04} / {\color{blue}0.9320} & {\color{blue}30.96} / {\color{blue}0.8528} & {\color{red}29.49} / {\color{blue}0.8148} & {\color{red}29.94} / {\color{blue}0.8832} & {\color{red}35.40} / {\color{blue}0.9542} \\
    \bottomrule
    \end{tabular}}
    \quad
    \caption{Quantitative comparison (average PSNR / SSIM) of NLSN~\cite{NLSN} and SwinIR~\cite{SwinIR} with different training datasets
    for $\times 4$ SR. ``+'' means the results are obtained with self-ensemble testing.}\label{tab:df2k_x4}
    \resizebox{\linewidth}{!}{
    \begin{tabular}{lcccccc}
    \toprule
    Method & Dataset & Set5 &  Set14 & B100 & Urban100 & Manga109 \\
    \midrule
    NLSN* & DIV2K  & 32.75 / 0.9018 & 28.96 / 0.7917 & 27.85 / 0.7464  & 27.24 / 0.8182 & 31.73 / 0.9225 \\
    NLSN*+  & DIV2K& {\color{blue}32.94} / 0.9036 & 29.10 / 0.7936 & 27.92 / 0.7482 & 27.47 / 0.8230 & 32.10 / 0.9253\\
    \midrule
    SwinIR+ & DF2K & {32.93} / {\color{red}0.9043} & {\color{blue}29.15} / {\color{red}0.7958} & {\color{blue}27.95} / {\color{red}0.7494} & {\color{blue}27.56} / {\color{blue}0.8273} & {\color{blue}32.22} / {\color{red}0.9273} \\
    \hdashline
    NLSN* & DF2K & 32.86 / 0.9025 & 29.11 / 0.7940 & 27.92 / 0.7481 & 27.49 / 0.8247 & 32.09 / 0.9251 \\
    NLSN*+ & DF2K  & {\color{red}32.99} / {\color{blue}0.9037} & {\color{red}29.19} / {\color{blue}0.7952} & {\color{red}27.97} / {\color{blue}0.7490} & {\color{red}27.66} / {\color{red}0.8279} & {\color{red}32.34} / {\color{blue}0.9269} \\
    \bottomrule
    \end{tabular}}
\end{table}
In Table~\ref{tab:df2k_x2}, Table~\ref{tab:df2k_x3} and Table~\ref{tab:df2k_x4}, we include the evaluation results of NLSN* trained on
DIV2K+Fllickr2K (DF2K) dataset for $\times 2$, $\times 3$ and $\times 4$ SR, respectively. As we can see, the performance of NLSN* can 
be further improved when trained on DF2K dataset, which shows that the proposed normalization strategy can also work with large training 
dataset with variety of texture details.

\section*{A3. Self-Ensemble Results}
It is well known that the performance of SR network could be greatly improved with self-ensemble strategy. To show this, we also include the 
self-ensemble results of NLSN* in Table~\ref{tab:df2k_x2}, Table~\ref{tab:df2k_x3} and Table~\ref{tab:df2k_x4} for $\times 2$, $\times 3$
and $\times 4$ SR, respectively. The models with self-ensemble strategy are marked by ``+''. Both NLSN* models trained on DIV2K dataset and 
DF2K dataset can get substantial performance improvements with self-ensemble testing strategy.
\begin{figure*}
    \centering
    \includegraphics[width=0.46\linewidth]{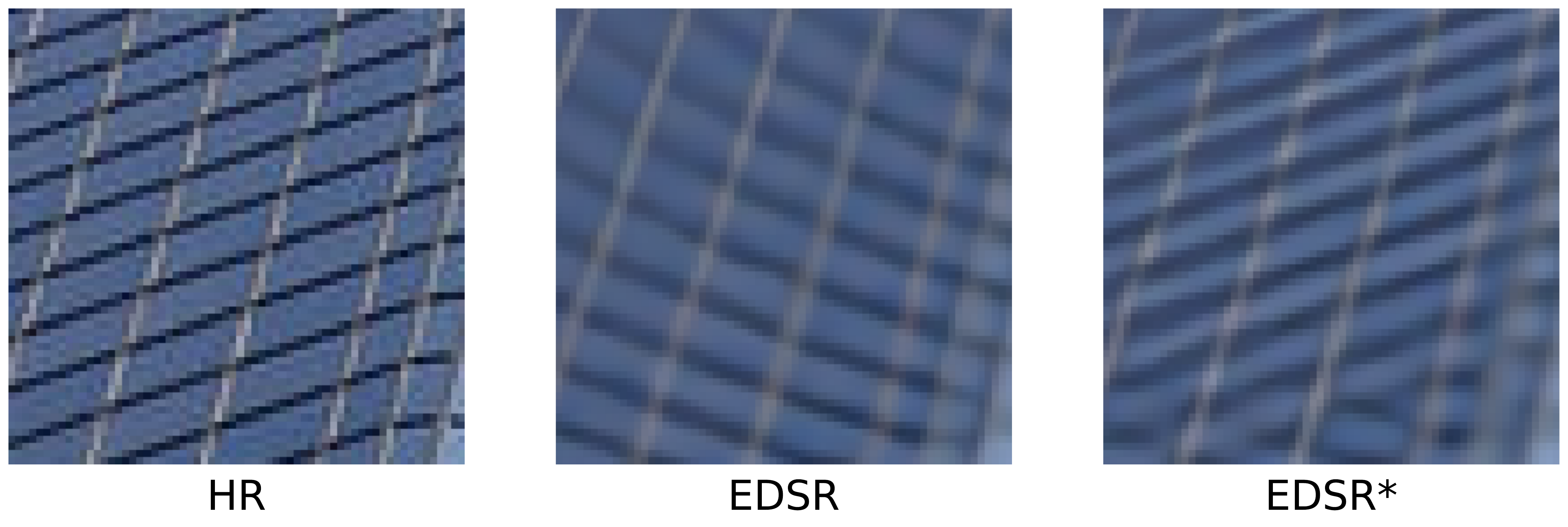}
    \includegraphics[width=0.46\linewidth]{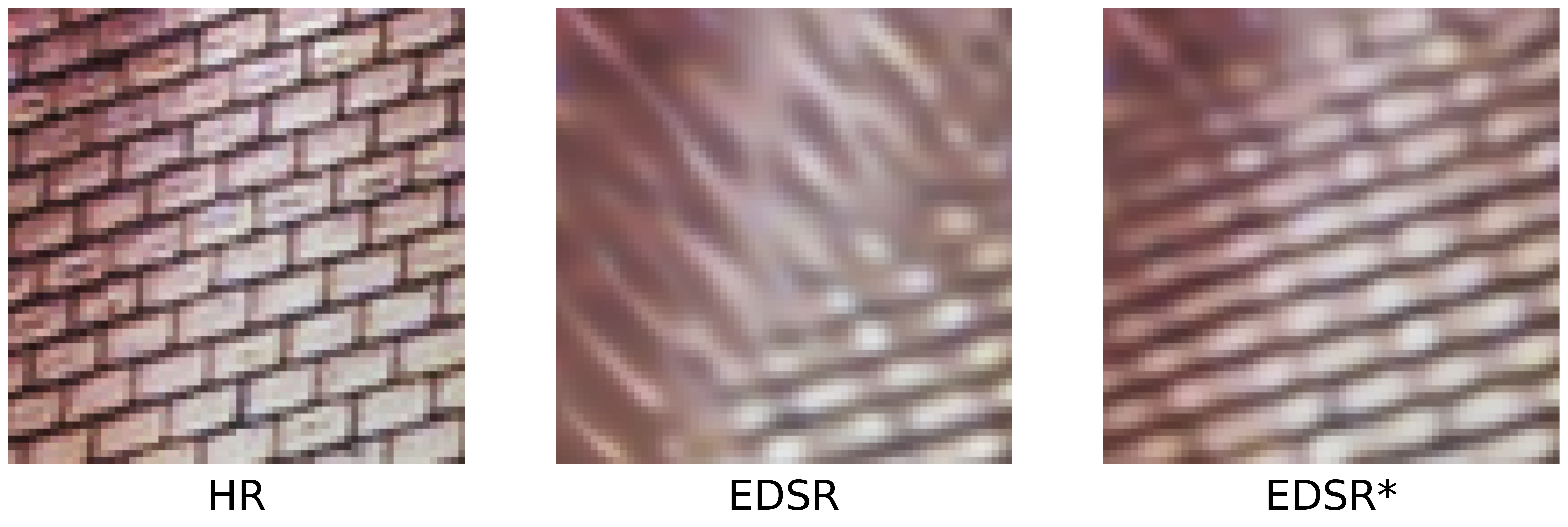}
    \includegraphics[width=0.46\linewidth]{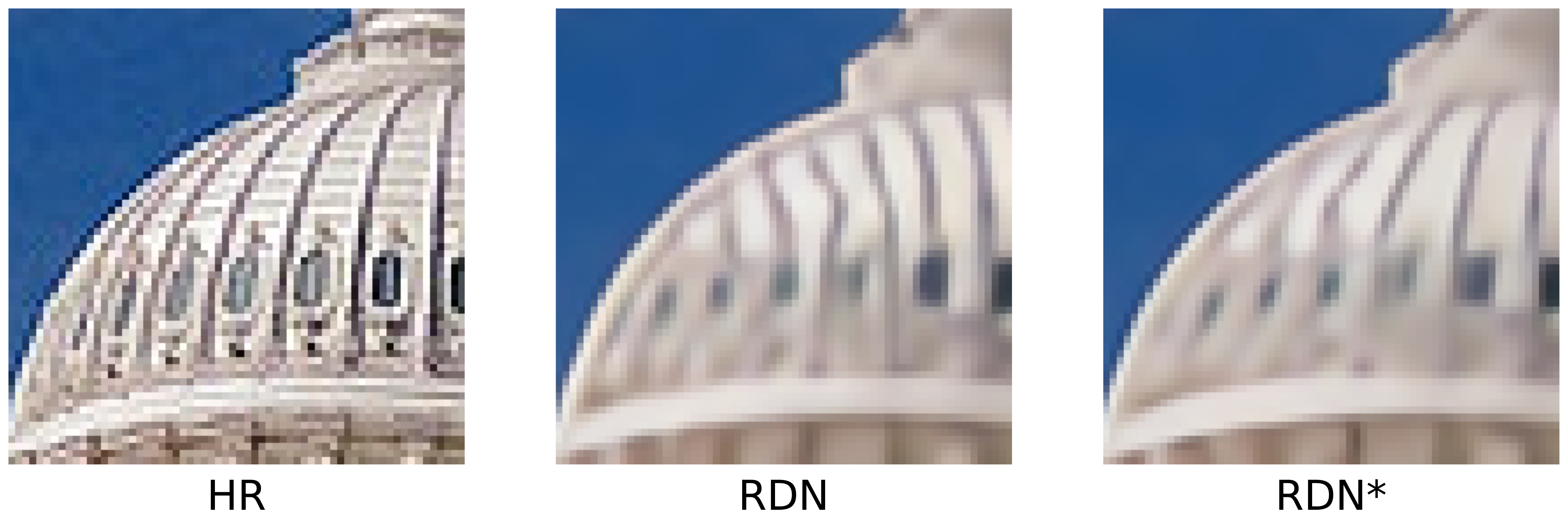}
    \includegraphics[width=0.46\linewidth]{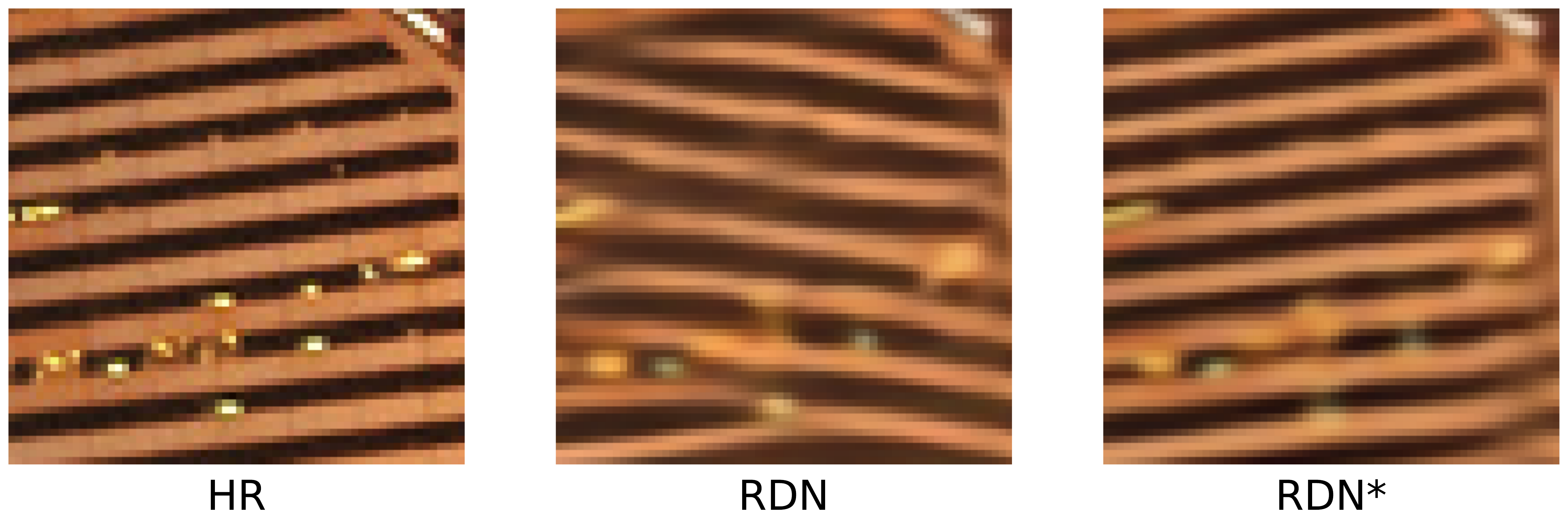}
    \includegraphics[width=0.46\linewidth]{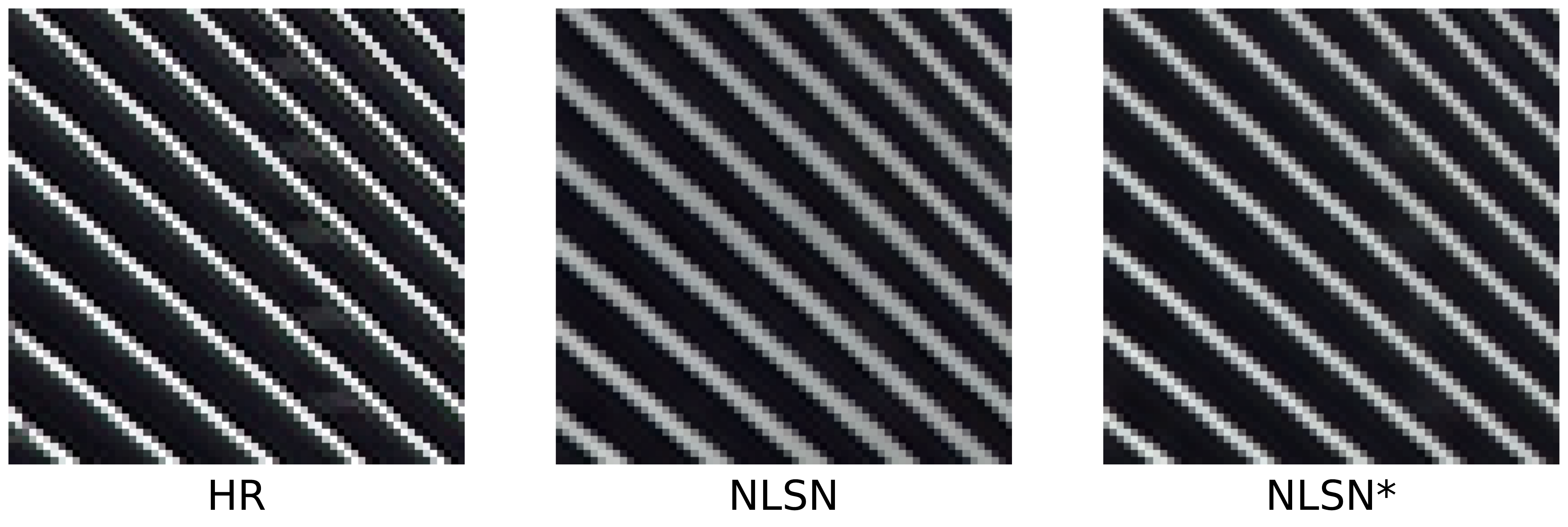}
    \includegraphics[width=0.46\linewidth]{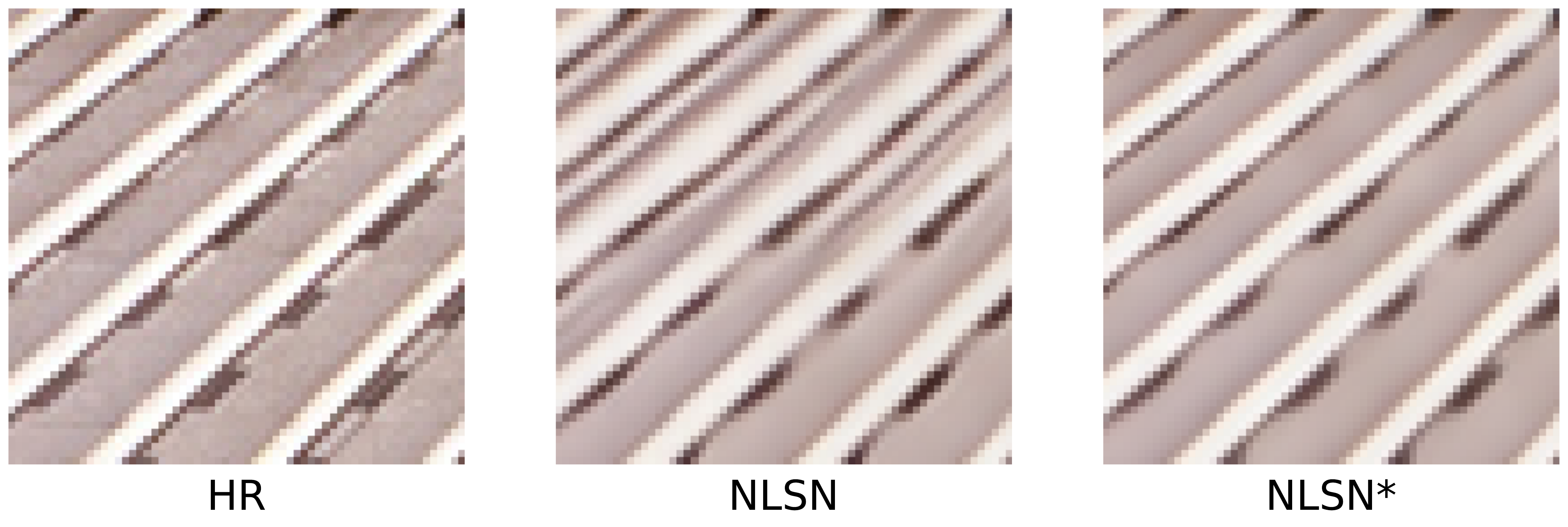}
    \includegraphics[width=0.46\linewidth]{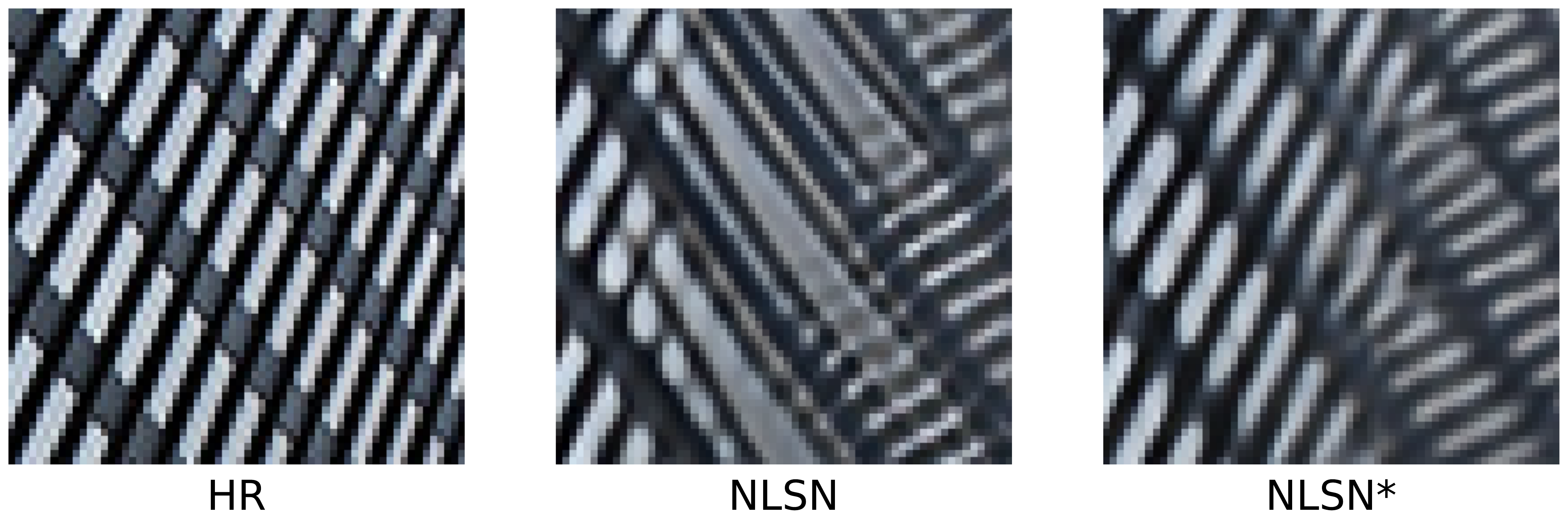}
    \includegraphics[width=0.46\linewidth]{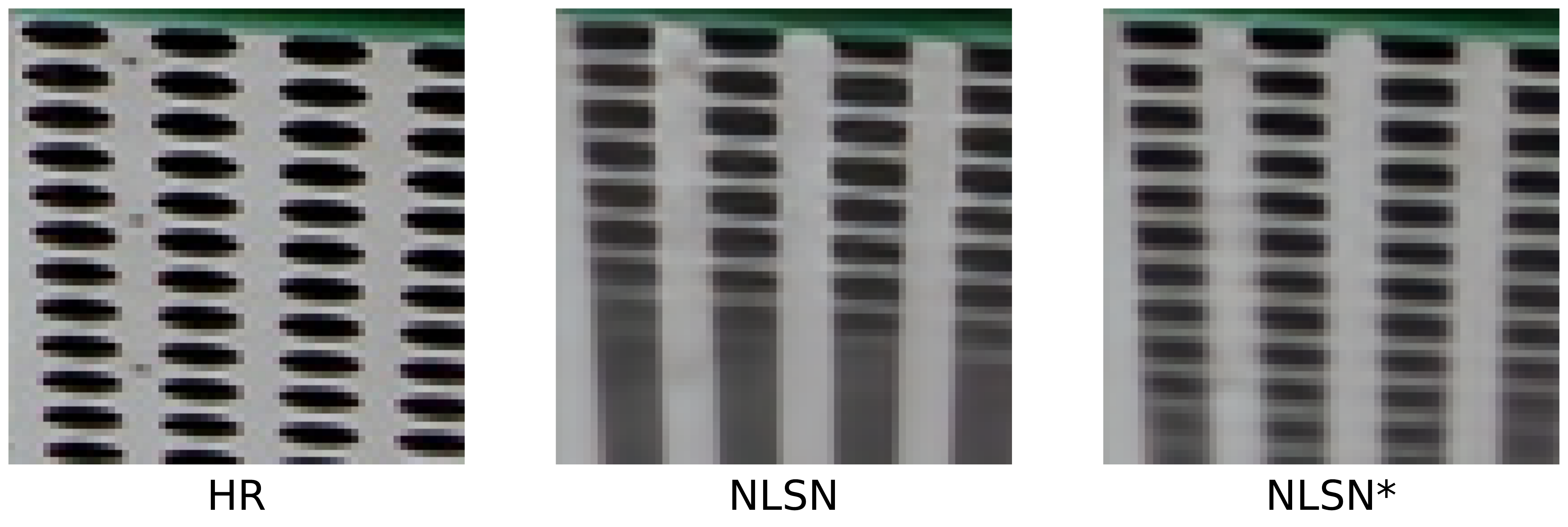}
    \caption{More visual results on Urban100 $\times 4$ dataset}\label{fig:more_vis}
\end{figure*}

\section*{A4. Comparison with SwinIR}
Recently, Transformer-based~\cite{Transformer} models become the leading methods for various of computer vision tasks. In image SR, the Swin Transformer~\cite{Swin} based
SwinIR~\cite{SwinIR} network achieved significant performance improvements compared with traditional CNN-based models. We also compare the NLSN* with SwinIR
in Table~\ref{tab:df2k_x2}, Table~\ref{tab:df2k_x3} and Table~\ref{tab:df2k_x4}. Both NLSN*+ and SwinIR+ are trained on DF2K dataset.
It can be observed that NLSN*+ can achieve comparable results with SwinIR+ on benchmark datasets, which shows the effectiveness of our AdaDM
in boosting the SR performance. With our AdaDM, the CNN-based NLSN~\cite{NLSN} network can even match the performance of Transformer-based SwinIR network.

\section*{A5. More Visual Results}
Our AdaDM can reconstruct the edges in images very well. To verify this, we show more visual comparison results in
Figure~\ref{fig:more_vis}. As we can see, the models with our AdaDM can recover the edges much better than
original models.

{\small
\bibliographystyle{ieee_fullname}
\bibliography{egbib}
}

\end{document}